\documentclass[nomathfonts]{aipproc}
\layoutstyle{6s}

\usepackage{xspace}
\usepackage{amsmath,amssymb}
\usepackage{graphicx}
\usepackage{multirow}

\graphicspath{{./EPS/}}

\input{alphabet}

\def\babs{\begin{abstract}}             \def\eabs{\end{abstract}}
\def\barr{\begin{array}}                \def\earr{\end{array}}
\def\bcc{\begin{center}}                \def\ecc{\end{center}}

\def\bdes{\begin{description}}          \def\edes{\end{description}}
\def\bdoc{\begin{document}}             \def\edoc{\end{document}}
\def\ben{\begin{enumerate}}             \def\een{\end{enumerate}}
\def\beqn{\begin{eqnarray}}             \def\eeqn{\end{eqnarray}}
\def\beqnl#1{\beqn\label{#1}}           \def\eeqnl#1{\label{#1}\eeqn}
\def\beqnx{\begin{eqnarray*}}           \def\eeqnx{\end{eqnarray*}}
\def\bseqn{\begin{subeqnarray}}         \def\eseqn{\end{subeqnarray}}
\def\beq#1\eeq{\begin{equation}#1\end{equation}}
\def\bal#1\eal{\begin{align}#1\end{align}}
\def\balx#1\ealx{\begin{align*}#1\end{align*}}
\def\beqx{$$}                           \def\eeqx{$$}
\def\bfig{\protect\begin{figure}}       \def\efig{\protect\end{figure}}
\def\bfigx{\protect\begin{figure*}}     \def\efigx{\protect\end{figure*}}
\def\bfigt{\protect\begin{figurette}}   \def\efigt{\protect\end{figurette}}
\def\bfl{\begin{flushleft}}             \def\efl{\end{flushleft}}
\def\bfr{\begin{flushright}}            \def\efr{\end{flushright}}
\def\bit{\begin{itemize}}               \def\eit{\end{itemize}}
\def\bmi{\begin{minipage}}              \def\emi{\end{minipage}}
\def\bfmi{\begin{fminipage}}            \def\efmi{\end{fminipage}}
\def\bpic{\begin{picture}}              \def\epic{\end{picture}}
\def\bqu{\begin{quote}}                 \def\equ{\end{quote}}
\def\bqun{\begin{quotation}}            \def\equn{\end{quotation}}
\def\bsl{\begin{slide}}                 \def\esl{\end{slide}}
\def\btabb{\begin{tabbing}}             \def\etabb{\end{tabbing}}
\def\btabl{\begin{table}}               \def\etabl{\end{table}}
\def\btablx{\begin{table*}}             \def\etablx{\end{table*}}
\def\btab{\begin{tabular}} 
\def\btabu{\begin{tabular}}             \def\etabu{\end{tabular}}
\def\btabx{\begin{tabular*}}            \def\etabx{\end{tabular*}}
\def\bbib{\begin{thebibliography}{999}} \def\ebib{\end{thebibliography}}
\def\bver{\begin{verbatim}}             \def\ever{\end{verbatim}}
\def\bca{\begin{cases}}                  \def\eca{\end{cases}}

\def\XS{\xspace}

\def\epsb{\ensuremath{\epsilonb}\XS}
\def\theb{\ensuremath{\thetab}\XS}
\def\sig2{\ensuremath{\sigma^2}}
\def\Wbx{\wb{_\mathrm{x}^j}}
\def\Wbs{\wb_\mathrm{s}^j}
\def\Wbe{\wb_\epsilon^j}
\def\Ws{w{_\mathrm{s}^j}}
\def\Wx{w{_\mathrm{x}^j}}
\def\iRe{\Rb_\epsilon^{-1}}
\def\Re{\ensuremath{\Rb_\epsilon}}
\def\Rt{\Rb_\tau}
\def\iRa{\Rb_a^{-1}}
\def\nWs{\|w_\textrm{s}^j\|^2}

\def\prior{\textit{prior}\XS}
\def\post{\textit{posterior}\XS}
\def\scbf#1#2{\noindent{\bf{\large #1}#2}\linebreak}

\def\reff#1{(\ref{#1})}

\def\pth#1{\left(#1\right)}
\def\Pth#1#2{\left(#1\big|#2\right)}
\def\Exp#1{\Ec\textrm{xp}\pth{#1}}
\def\IG#1#2{\Ic\Gc\Pth{#1}{#2}}

\def\sca#1{\textsc{#1}}

\title{Bayesian Wavelet Based Signal and Image Separation}
\author{Mahieddine M. Ichir}
	{address = {Laboratoire des Signaux et Syst\`emes,\linebreak
  	Sup\'elec, Plateau de Moulon, 3 rue Joliôt Curie, 91192 Gif-sur-Yvette, France},
  	email = {ichir@lss.supelec.fr}}
\author{Ali Mohammad-Djafari}
	{address = {Laboratoire des Signaux et Syst\`emes,\linebreak
  	Sup\'elec, Plateau de Moulon, 3 rue Joliôt Curie, 91192 Gif-sur-Yvette, France},
  	email = {djafari@lss.supelec.fr}}

\begin{document}

\begin{abstract}
In this contribution, we consider the problem of blind source separation in a
Bayesian estimation framework. The wavelet representation allows us to assign an adequate \prior distribution to the wavelet coefficients of the sources. MCMC algorithms are implemented to test the validity of the proposed approach, and the non linear approximation of the wavelet transform is exploited to aleviate the algorithm.
\end{abstract}

\maketitle

\section{Introduction}
We find applications of blind source separation (BSS) in many fields of data analysis: chemistry, medical imaging (EEG, MEG), seismic data analysis and astronomical imaging. Many solutions have been developped to try to solve this problem: Independant Component Analysis (ICA) \cite{Comon94a,Hyvarinen01a}, maximum likelihood estimation \cite{Gaeta90a}, and methods based on second or higher order statistics of the signals \cite{Belouchrani97a,Cardoso99a}. These methods have proved their efficiency in many applications, however they do not apply for noisy observations models.

A different approach has been considered to solve the BSS problem, we find in \cite{Rowe98a,Djafari99d,Knuth99} an introductory analysis of the problem in a Bayesian estimation framework. Some of the methods outlined earlier can be reformulated via the Bayes rule, and a similar formalism can be obtained.

In this contribution, we treat the BSS problem in a Bayesian estimation framework. As in previous works on this subject \cite{Djafari02f,Ichir03a}, the problem is transported to a transform domain: the wavelet domain. The advantage of such an approach is that some invertible transforms restructure the data, leaving them structures simpler to model, and this, as will be seen later, is useful in the formulation of the problem as an inference problem.

The paper is organized as follows: In section-II we present the BSS problem, write the associated equations and introduce the Bayesian solution of the problem. In section-III, we transport the problem to a transformed data space (wavelet) and give the justification for that approach. In section-IV, we present the associated MCMC-based optimization algorithm. We consider then the non-linear approximation of the wavelet transform to introduce a denoising procedure by some thresholding rule. At the end, we conclude and give future perspectives of the present work.

\section{Bayesian blind source separation (BBSS)}
Blind source separation (BSS) consists of recovering unobservable sources from a set of their instantaneous and linear mixtures. The direct observational model can be described by:
\beq
\xb(t) = \Ab\sb(t) + \epsb(t),\quad t\in \Cc
\label{BSS.mod}
\eeq
where $\Cc = \{\eZ\textrm{ : time series signals},\eZ^2 \textrm{ : 2D images}\}$, $\xb(t)_{t=1,...,T}$ is the observed $m$-column vector, $\sb(t)_{t=1,...,T}$ is the unknown sources $n$-column vector, \Ab is the $(m\times n)$ full rank matrix, and $\epsb(t)_{t=1,...,T}$ is the noise $m$-column  vector.

The Bayesian approach to BSS starts by writing the \post distribution of the sources, jointly with the mixing matrix and any other parameters needed to describe the problem:
\beq
P\Pth{\sb,\Ab,\Re}{\xb} \propto P\Pth{\xb}{\sb,\Ab,\Re}\pi\pth{\sb,\Ab,\Re}
\label{bay.bss}
\eeq
where $P\Pth{\sb,\Ab,\Re}{\xb}$ is the joint \post distribution of the unknown sources, the mixing matrix and the noise covariance matrix. $P\Pth{\xb}{\sb,\Ab,\Re}$ is the likelihood of the observed data $\xb$ and $\pi\pth{\sb,\Ab,\Re}$ is the \prior distribution that should reflect the prior information we may have about \sb, \Ab and \Re. The noise $\epsilonb(t)$ is assumed Gaussian, spacially independant and temporarily white: $\Nc\pth{0,\Re}$, with $\Re = \textrm{diag}\pth{\sig2_1,...,\sig2_m}$.

An important step in Bayesian inference problems is to assign appropriate expressions to $\pi\pth{\sb,\Ab,\Re}$. The likelihood $P\Pth{\xb}{\sb,\Ab,\Re}$ is determined by the hypotheses made on the noise $\epsb(t)$. It is reasonable to assume that the sources, the mixing matrix and the noise variance are independant:
\beq
\pi\pth{\sb,\Ab,\Re} = \pi\pth{\sb}\pi\pth{\Ab}\pi\pth{\Re}
\eeq
The \prior distribution $\pi\pth{\Ab}$ can be determined by some physical knowledge of the mixing mechanism. In our work, the mixing matrix is assigned a Gaussian prior distribution:
\beq
\pi\pth{\Ab} = \prod_{i,j}\Nc\Pth{a_{ij}}{\mu_{ij},\sig2_{ij}}
\eeq
The appropriate selection of prior distributions is still a subject of intensive research. We find in \cite{Snoussi02c,Rodriguez03a} some interesting work on this topic. We thus define, as results of these works, a Gamma prior distribution for the inverse of the variances:
\beq
\pi\pth{x} = \Gc({x}|{\alpha,\theta}) \propto {x^{\alpha-1}}{e^{-\frac{x}{\theta}}}
\eeq

Some work has been done on BSS \cite{Snoussi02b,Snoussi02a}, by assigning a mixture of Gaussians prior to the sources:
\beq
\pi\pth{s} = \sum_{l=1}^{L}p_l~\Nc_l\Pth{s}{\mu_l,\tau_l}, \quad \sum_{l=1}^{L} p_l = 1
\label{mix.gau}
\eeq
This distribution is very interesting, any distribution $\pi^*\pth{s}$ can be well aproximated by a Gaussian mixture distribution, and the higher $L$ (number of Gaussians), the better the approximation, but the higher the complexity of the associated model. The difficulty lies then in \textit{how $L$} \textit{should be chosen to well approximate the distribution with reasonable complexity ?}. We note that for $L$ Gaussians, we need $(3L-1)$ parameters $(p,\mu,\tau)$ to totally define the mixture.

\section{BBSS in the wavelet domain}
An idea, that has been exploited with success, is to treat the problem in a tranform domain. We find in \cite{Snoussi01e} a proposed solution to a spectral BSS problem. In \cite{Djafari02f,Ichir03a}, a first approach to the problem has been treated in the wavelet domain. The particular properties of these transforms: \textit{linearity} and \textit{inversibility} makes that the BSS problem is formulated in a similar manner and that we can go back and forth without any difficulty. The BSS problem described by equation \reff{BSS.mod} is rewritten in the wavelet domain as:
\beq
\Wbx(k) = \Ab\Wbs(k) + \Wbe(k) \quad k\in\Cc, j = 1,\ldots, J
\label{BSS.mod.wav}
\eeq
where $\Cc = \{\eZ\textrm{ : time series signals},\eZ^2 \textrm{ : 2D images}\}$, and :
\beq
\Ws(k) = ~<s(t),\psi^j(t-k)> ~= \int_\Cc s(t)~\psi^j(t-k)~ dt
\eeq
where $\psi^j(t)=2^{-j/2}\psi\pth{2^{-j}{t}}$. We point out to the fact that the statistical properties of the noise does not change in the wavelet doamin:
\beq
\epsilon(t) \sim \Nc\pth{0,\sigma_\epsilon^2} \Longrightarrow w_\epsilon^j(k) \sim \Nc\pth{0,\sigma_\epsilon^2}
\eeq
We will refer by $\Wbs(k), \Wbx(k)$ and $\Wbe(k)$ to the wavelet coefficients vectors of $\sb(t), \xb(t)$ and $\epsb(t)$ at resolution $j$, respectively. The $k$-index will be dropped to aleviate the expressions since $\Wbs$ and $\Wbe$ are temporarily white, and thus $\Wbs(k)$ and $\Wbs$ define identically the same vector unless specified.\\
The \post distribution of the \textit{new unknowns} is now given by:
\beq
P\Pth{\Wbs,\Ab,\Re}{\Wbx}
\propto P\Pth{\Wbx}{\Wbs,\Ab,\Re}\pi\pth{\Wbs}\pi\pth{\Ab}\pi\pth{\Re}
\label{post.equ.wav}
\eeq
The wavelet transform has some particular properties that make it interesting for Bayesian formulation of the BSS problem:
\begin{description}
\item[locality] each wavelet atom $\psi_j(t-k)$ is localised in time and frequency.
\item[edge detection] a wavelet coefficient is significant if and only if an irregularity is present within the support of the wavelet atom.
\end{description}

These two properties have a great impact on the wavelet (1D/2D)-statistical signal processing. The wavelet coefficients can be reasonably considered \textit{uncorrelated} due to \textit{locality} (we say that the wavelet transform acts as decorrelator), and assigned a separable probabilty distribution:
\beq
\pi\Big\{\bigcup_{j,k}\Ws(k)\Big\} = \prod_{j,k}\pi\pth{\Ws(k)}
\eeq

The second property (\textit{edge detection}) has a consequence on the type of the distribution we will assign to the wavelet coefficients:

\begin{center}
\parbox{.9\textwidth}{\it The wavelet transform of \textit{natural} sources results in a \textbf{large} number of \textbf{small} coefficients, and a \textbf{small} number of \textbf{large} coefficients.}
\end{center}

\begin{figure}[!htb]
\begin{tabular}{cc}
\includegraphics[height=45mm]{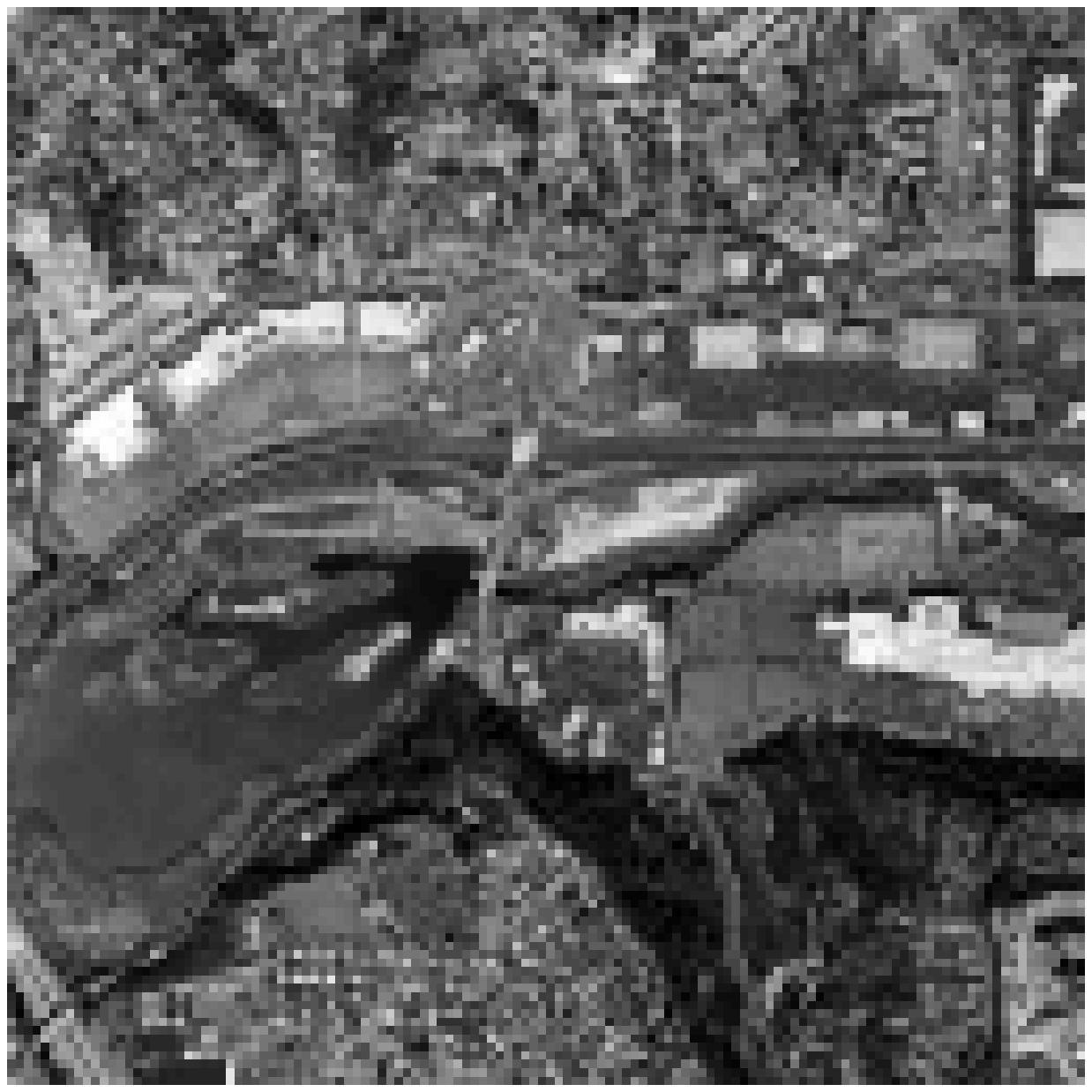} &
\includegraphics[height=45mm]{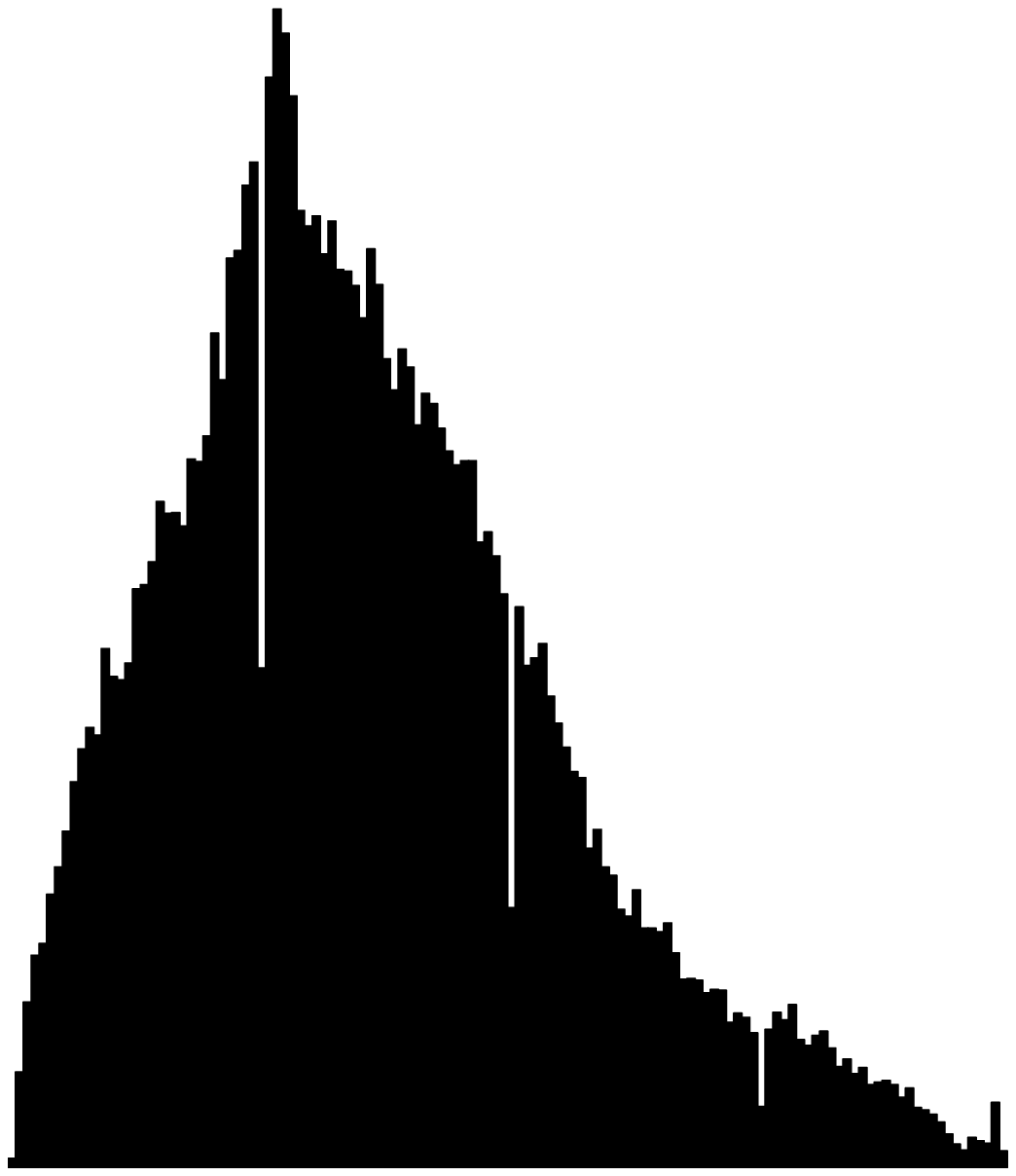} \\
a. & b.\\
\includegraphics[height=45mm]{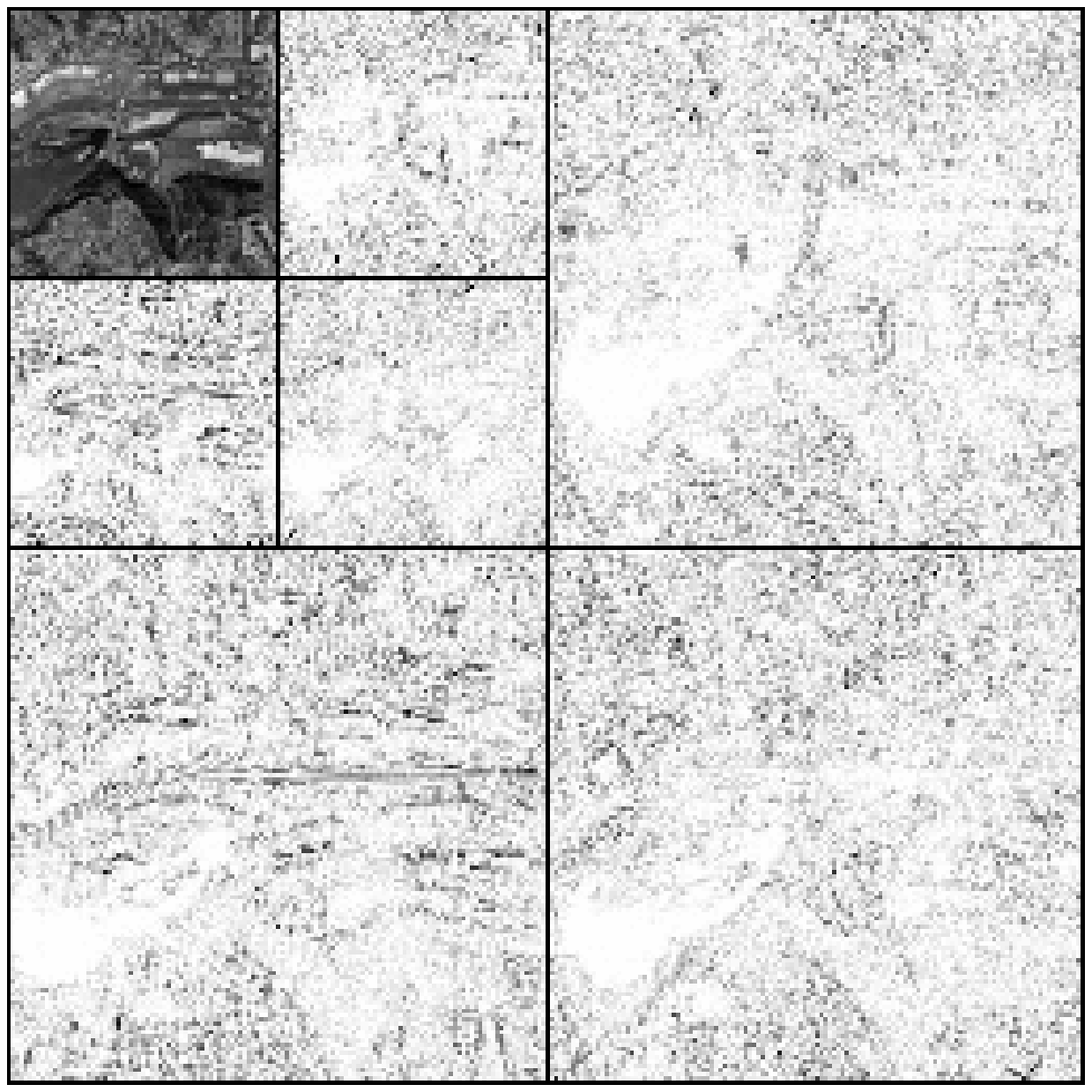} &
\includegraphics[height=45mm]{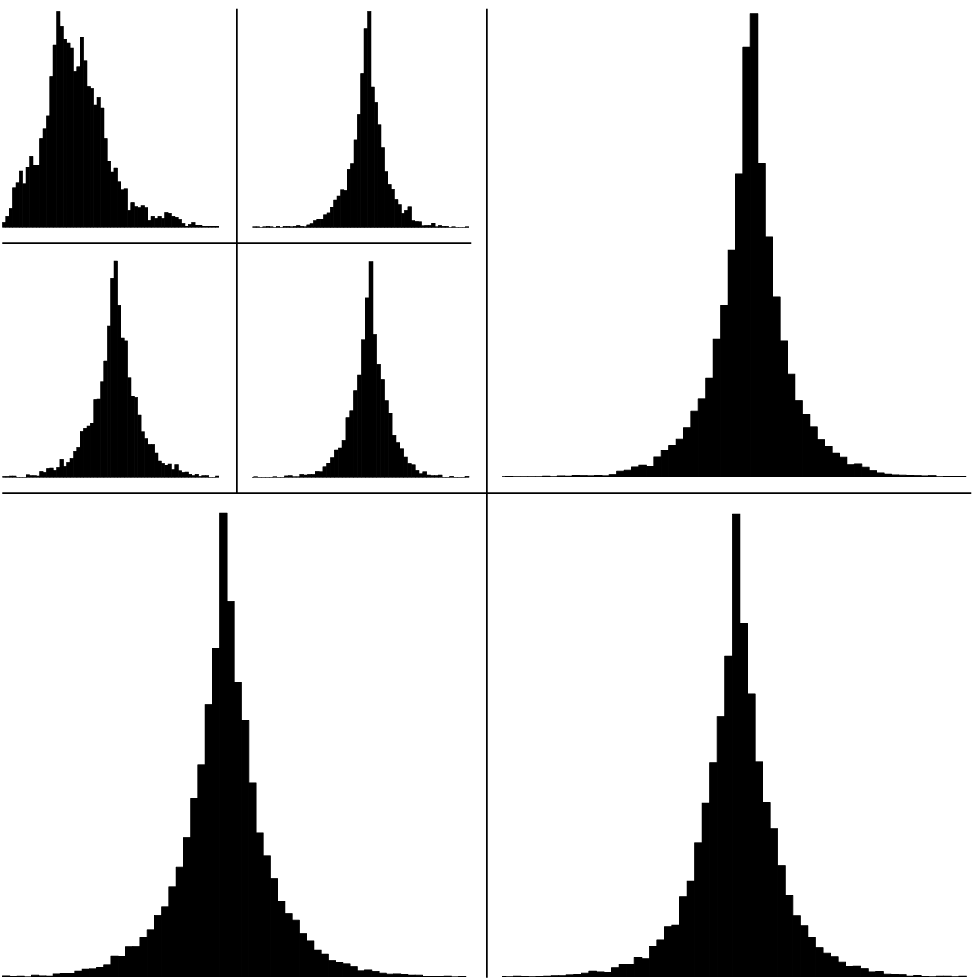} \\
c. & d.
\end{tabular}
\caption{Sparsity property of the wavelet coefficients: a. aerial image, b. histogramme of image (a), c. the wavelet transform of image (a), d. histograms of the wavelet coefficients in the different bands (c)}
\label{Fig.Spars}
\end{figure}

This property (sparsity) is shown in Figure \reff{Fig.Spars}. The \prior distribution of the wavelet coefficients is then very well approximated by centered, peaky and heavy tailed like distributions. Mallat has porposed in \cite{Mallat89a} to model the wavelet coefficients by generalized exponential distributions:
\beq
P(.) = K \Exp{-\frac{1}{\gamma}|.|^\alpha}, \quad \gamma >0, 1\leq\alpha\leq 2
\label{Exp.equ}
\eeq
Crouse in \cite{Crouse98a} has assigned to the wavelet coefficients a Gaussian mixture distribution to capture the sparsity characteristic:
\beq
P(.) = p~ \Nc\Pth{.}{0,\tau_L} + (1-p)~ \Nc\Pth{.}{0,\tau_H}, \quad \tau_H >> \tau_L
\label{GM.equ}
\eeq
where $p =$ Prob.(wavelet coefficient $\in$ low energy state). In the sequel, we will only emphasize on the Gaussian mixture model. For the generalized exponential case, we refer to \cite{Ichir03a}. Note that we choose a two Gaussian mixture model with a total number of parameters equals to three. 

\section{MCMC implementation}
\label{Implementation}
Once we have defined the priors and properly written the \post distribution $P\Pth{\Wbs,\Ab,\Re}{\Wbx}$, we define a \post estimates of the different parameters that characterizes the BSS problem. To do this, we will generate samples from the joint distribution \reff{post.equ.wav}, by means of MCMC algorithms (Monte Carlo Markov Chain methods) and than choose the \post means as estimates.

\subsection*{Hidden variables}
The conditional \post distribution of the sources coefficients is a mixture of Gaussians of the type:
\beq
P\Pth{\Wbs}{\Wbx,\Ab,\Re}
\propto \Nc\Pth{\Wbx}{\Ab\Wbs,\Re}\pi\Pth{\Wbs}{\theta}
\eeq
where 
$$\displaystyle\pi\Pth{\Wbs}{\theta} = \prod_i^n\pi\Pth{{\Ws}_i}{\theta} = \prod_i^n\sum_l^Lp_{li}\Nc\Pth{{\Ws}_i}{0,\tau_{li}}$$
where $i$ stands for the $i$-th source. The complexity of such model increases with increasing $n$ (for a 2-Gaussians wavelet model, a total of $(2L-1)^n = 3^n$ parameters has to be defined in order to describe the model). Thus the introduction of a label variable $\zb^j \in \{1,\ldots,L\}^n = \{1,2\}^n = \{$Low state, High state$\}^n$ and a conditional parametrisation of the form:
\beq
\pi\Pth{\Ws}{\theta,z^j\in{[L,H]}} = \Nc\Pth{\Ws}{0,\tau_{[L,H]}}
\eeq
with $P(z^j\in L) = p_L$, and $P(z^j\in H) = p_H = 1-p_L$.

\subsection*{The MCMC Algorithm}

{\noindent}\underline{The hidden variables}
\begin{eqnarray*}
1.\quad\zb^j~\sim~P\Pth{\zb^j}{\Wbx,\theb} &=& \int_{w_s} P\Pth{\zb^j,\Wbs}{\Wbx,\theb} \\
&=&\pi\pth{\zb^j}\int_{w_s} \Nc\Pth{\Wbx}{\Ab\Wbs,\Re}\pi\Pth{\Wbs}{\zb^j,\Rt}
\end{eqnarray*}
where $\pi\Pth{\Wbs}{\zb^j,\Rt} = \Nc\Pth{\Wbs}{0,\Rt}$, and $\Rt = \textrm{diag}\pth{\tau_1,...,\tau_n}$.

{\noindent}\underline{The sources wavelet coefficients}
\begin{eqnarray*}
2.\quad\Wbs~\sim~P\Pth{\Wbs}{\Wbx,\zb^j,\theb} &=& \Nc\Pth{\Wbx}{\Ab\Wbs,\Re} \Nc\Pth{\Wbs}{0,\Rt}\\
&=& \Nc\Pth{\Wbs}{\mub_{s/\zb},\Rb_{s/\zb}}
\end{eqnarray*}
where $\mub_{s/\zb} = \Rb_{s/\zb}{\iRe\Ab^\dag\Wbx}$, and $\Rb_{s/\zb} = \pth{\Ab^\dag\iRe\Ab + \Rb_\tau^{-1}}^{-1}$.

{\noindent}\underline{The mixing matrix}
\begin{eqnarray*}
3.\quad\Ab~\sim~P\Pth{\Ab}{\Wbx,\theb} &=& \Nc\Pth{\Wbx}{\Ab\Wbs,\Re} \Nc\Pth{\Ab}{\mub_a,\Rb_a} \\
 & = & \Nc\Pth{\Ab}{\mub_A,\Rb_A}
\end{eqnarray*}
where ${vec}\footnote{$vec(.)$ is the vector representation of a matrix.}\pth{\mub_A} = \Rb_A \Big(\pth{\iRe\otimes\mathbb{I}_n} {vec}\pth{\Cb_{xs}} + \mub_a\Big)$, $\Rb_A = \pth{\iRe\otimes\Cb_{ss} + \iRa}$,\\
$\Cb_{ss} = \sum_{j,k} {\Wbs}{\Wbs}^\dag$ ~and~ $\Cb_{xs} = \sum_{j,k} {\Wbx}{\Wbs}^\dag$.

{\noindent}\underline{The hyperparameters}
\begin{equation*}
4.\quad\theb~\sim~P\Pth{\theb}{\Wbx,\Wbs,\Ab} = P\Pth{\Wbx}{\Wbs,\Ab,\theb}\pi\pth{\theb}\\
\end{equation*}
where $\theb$ stands for the the noise covariance matrix $\Re$ and the mixture parameters $\Rb_\tau = \textrm{diag}\pth{\tau_1,...,\tau_n}$ (variances of the Gaussians in the mixture).

{\noindent}\underline{The noise covariance}
\begin{eqnarray*}
4.a.\quad\sig2_{\epsilon_i}\sim~P\Pth{\sig2_{\epsilon_i}}{\Wbx,\Wbs,\Ab} &=&\Nc\Pth{\Wx_i}{[\Ab\Wbs]_i,\sig2_{\epsilon_i}}\IG{\sig2_{\epsilon_i}}{2,1} \\
&=&\IG{\sig2_{\epsilon_i}}{\alpha,\theta_i}, \quad i=1,...,m
\end{eqnarray*}
where $\alpha = T/2 + 2$ and  $1/\theta_i = \big(\sum_k \left(\Wx_i-[\Ab\Wbs]_i\right)^2/2 + 1 \big)$.

{\noindent}\underline{The Gaussians variances}
\begin{eqnarray*}
4.b.\quad\tau_i^j[L,H]~\sim~P\Pth{\tau_i^j}{w_{s_i}^j} &=&\Nc\Pth{w_{s_i}^j}{0,\tau_i^j}\IG{\tau_i^j}{2,1} \\
&=&\IG{\tau_i^j}{\alpha^j,\theta_i^j}, \quad i=1,...,n
\end{eqnarray*}
where $\alpha^j = T/2^{j} + 2$ and  $1/\theta_i^j = \big(\sum_{k}\big(w_{s_i}^j.\mathbb{I}_{(z_i^j=l)}\big)^2/2+1\big), \quad l = \{L,H\}$.

{\noindent}\underline{The prior probabilities}
\begin{eqnarray*}
5.\quad[p_{iL}^j,p_{iH}^j]~\sim~P\Pth{p_{iL}^j,p_{iH}^j}{\theta} &=&\Dc_2\pth{u_1+n_{iL},u_2+n_{iH}}, \quad i=1,...,n
\end{eqnarray*}
where $n_{il} = \sum_k \mathbb{I}_{(z^j_i = l)}$, and $\Dc_2\pth{\gamma_1,\gamma_2}$ stands for the Dirichlet distribution with parameters $(\gamma_1,\gamma_2)$ for the probability variables $(p_L,p_H=1-p_L)$.

\section{Simulation results}
\begin{figure}[!htb]
\begin{tabular}{cc}
a.\includegraphics[height=50mm]{OSource1} &
\includegraphics[height=50mm]{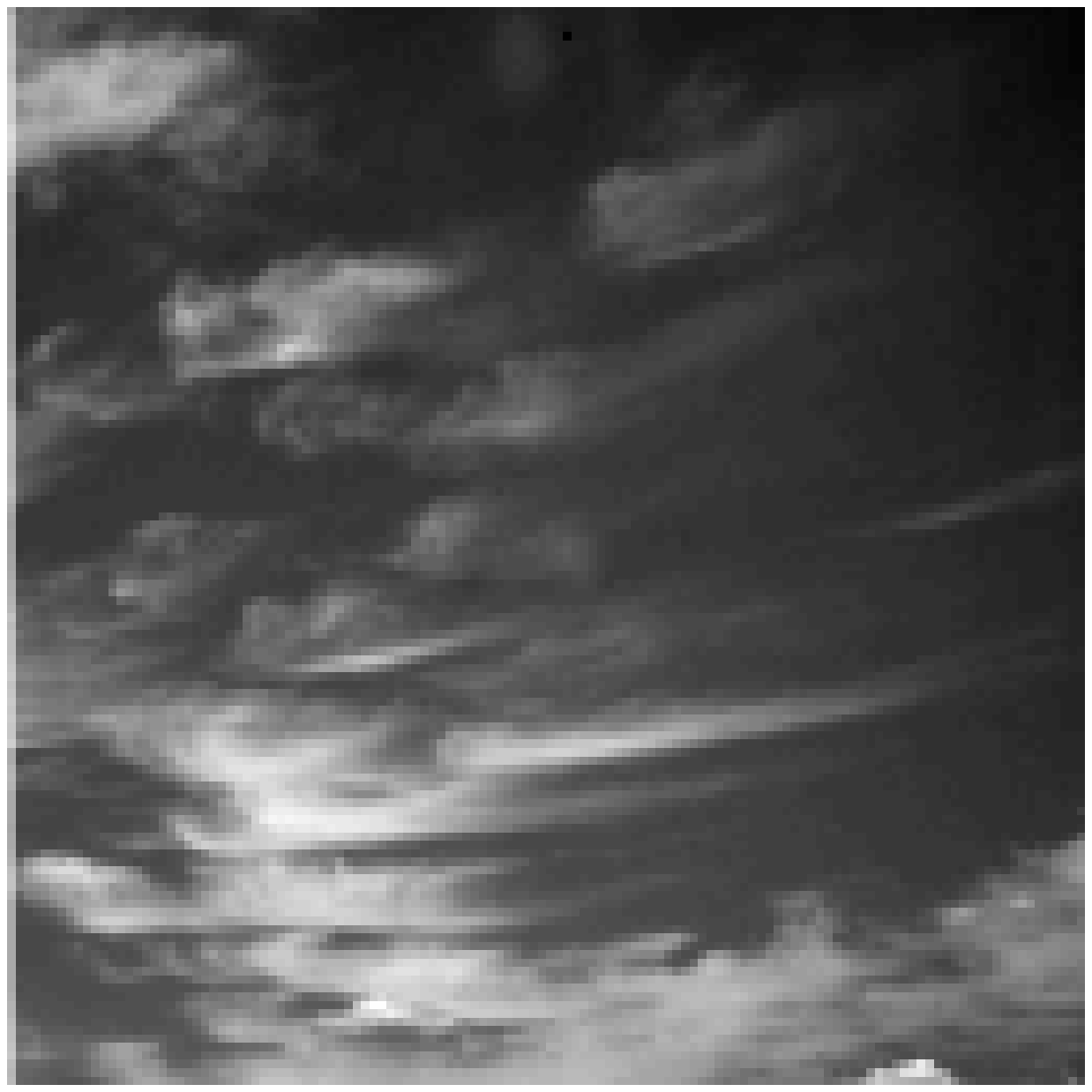} \\
b.\includegraphics[height=50mm]{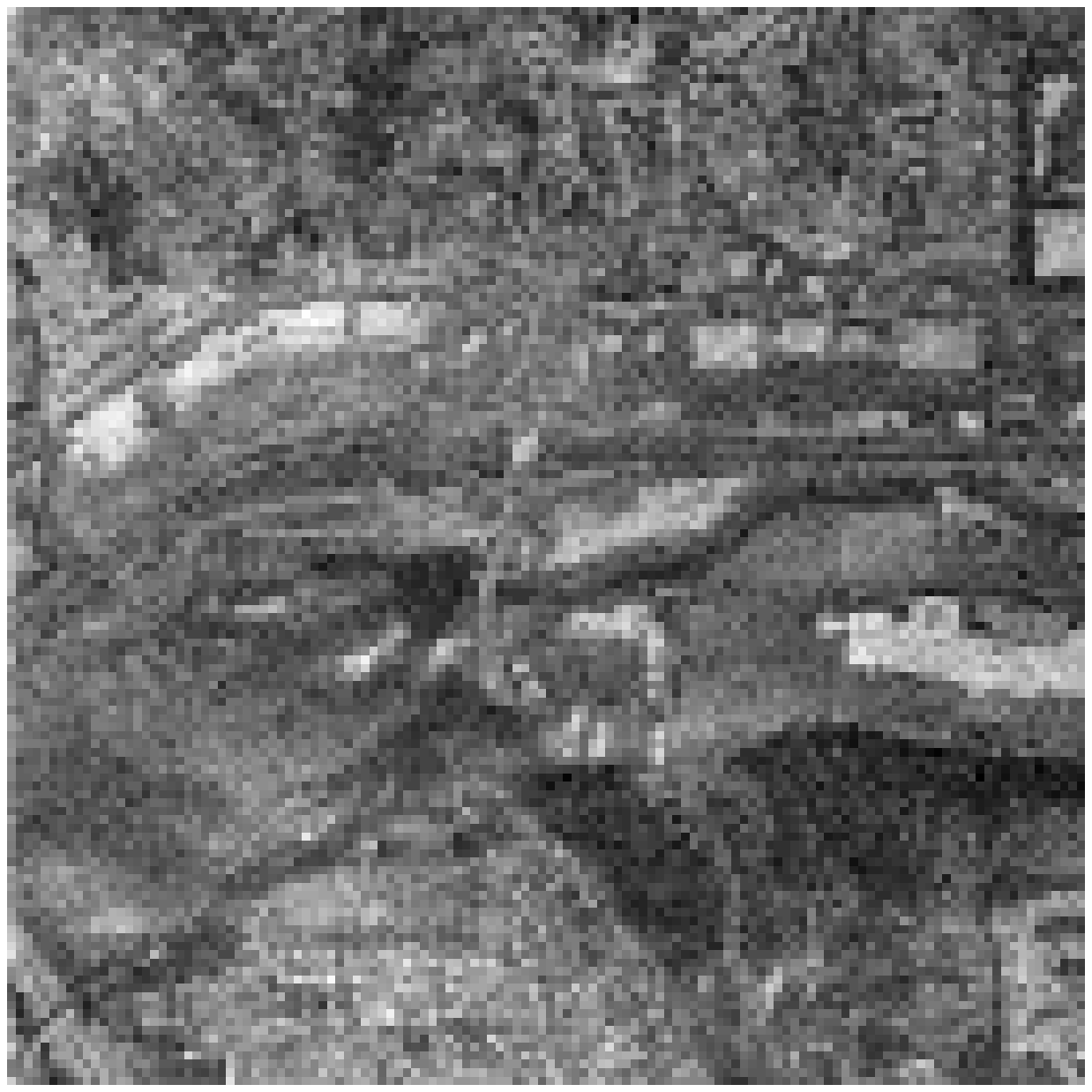} &
\includegraphics[height=50mm]{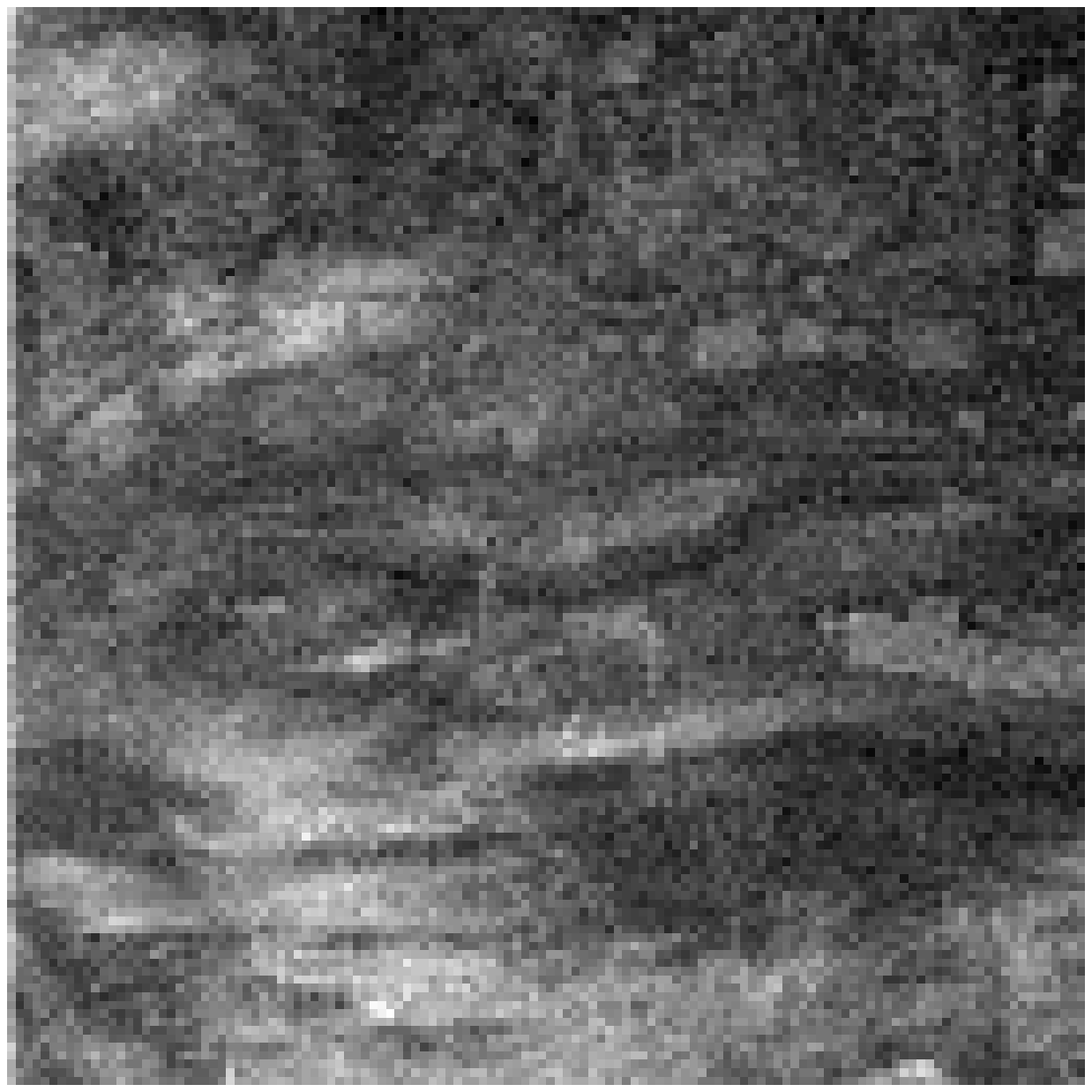} \\
$\delta = .16 $ & $\delta = .17 $ \\
c.\includegraphics[height=50mm]{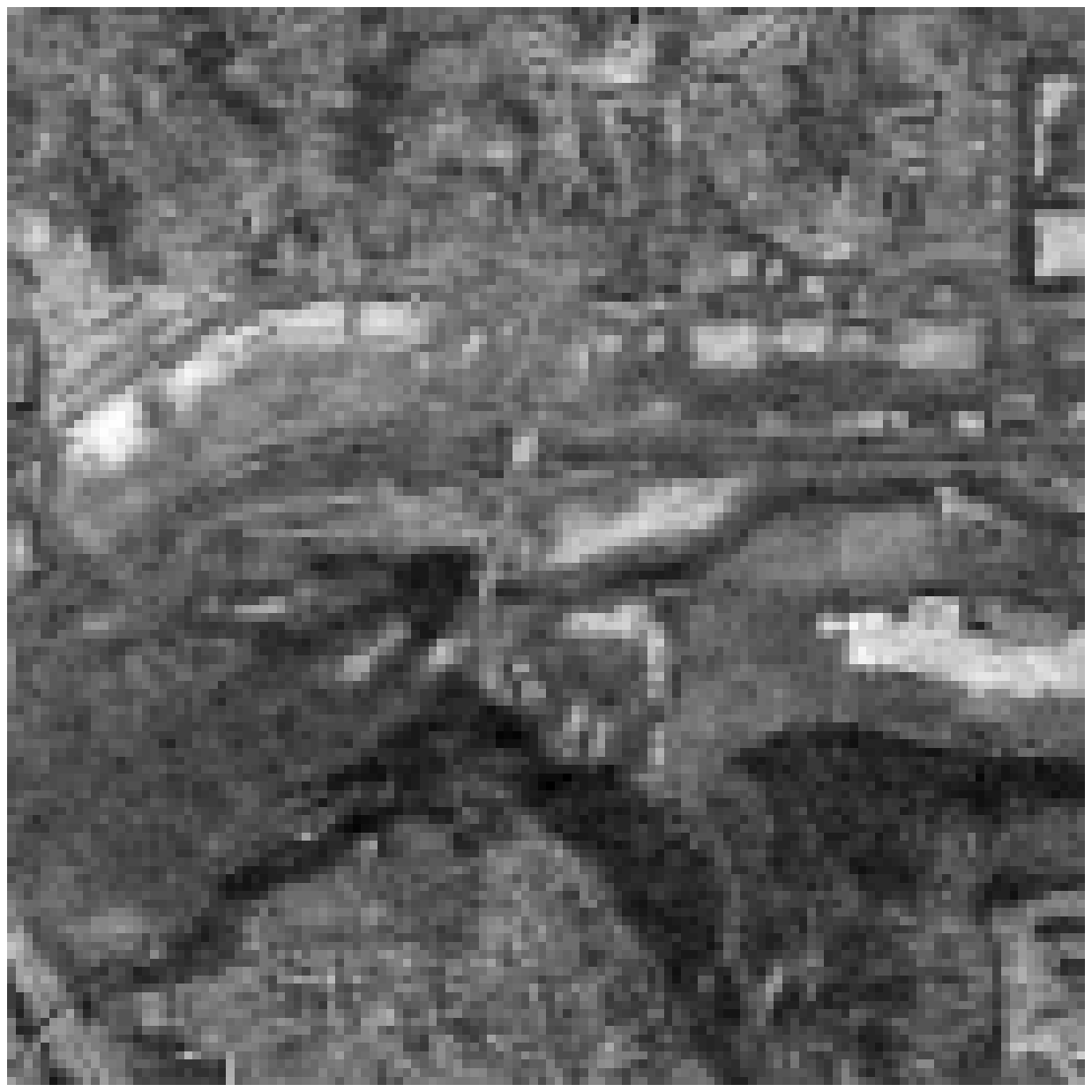} &
\includegraphics[height=50mm]{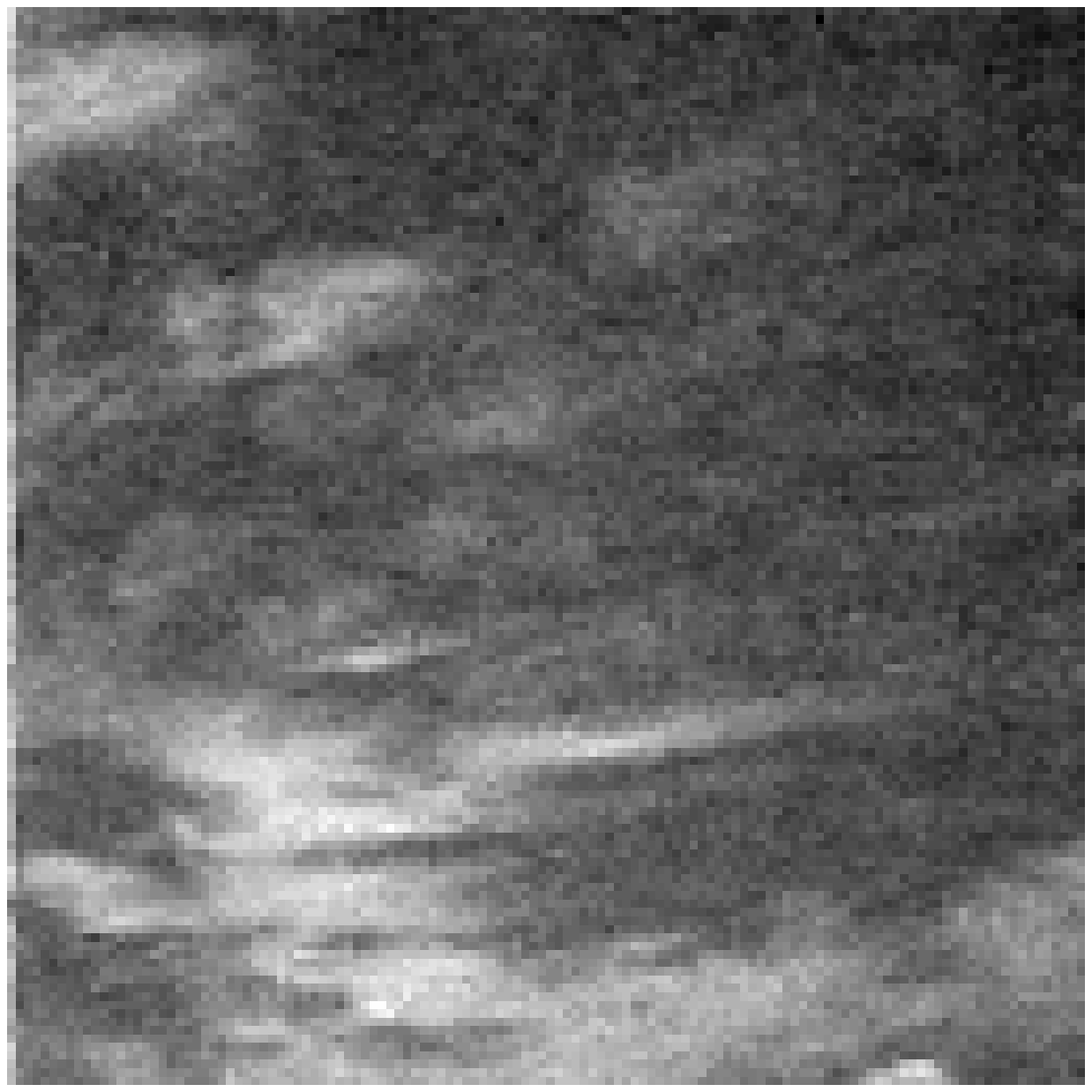} \\
$\delta = .05 $ & $\delta = .04 $ \\ 
\end{tabular}
\caption{a. original (128 $\times$ 128 pixels) sources, b. linearily mixed and noisy observations, c. estimated sources (MCMC algorithm)}
\label{Sml.Data}
\end{figure}

\begin{figure}[!htb]
\includegraphics[width=130mm,height=60mm]{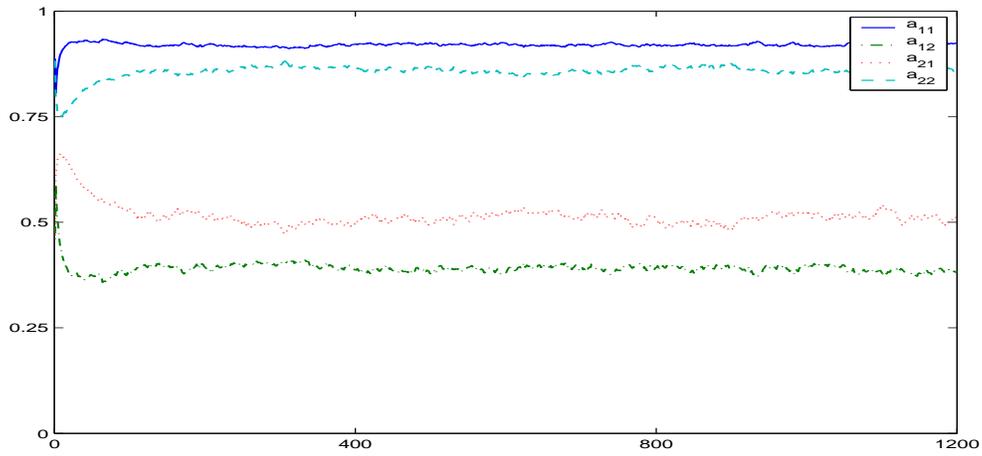}
\caption{evolution of the estimation of the elements of the matrix $\Ab$ during the iterations}
\label{Con.A}
\end{figure}

To verify the plausibilty of the proposed algorithm, we have made some tests on simulated data (128 x 128 pixels). In figure \ref{Sml.Data}.a, we present an aerial image and a cloud image that were linearily mixed to obtain the observed data in figure \ref{Sml.Data}.b. The mixing matrix is of the form:
\[
\Ab = \left[
\begin{array}{cc}
.91 & .49 \\
.42 & .87
\end{array}
\right]
\]
The signal to noise ratio is of $20$dB. The Symmlet-4 wavelet basis has been chosen (with 4 vanishing moments). The obtained estimates of the sources are presented in figure \ref{Sml.Data}.c. The evolution of the estimates of the elements of the matrix is presented in figure \ref{Con.A}, where the empirical posterior mean is found to be:
\[
\hat{\Ab} = \left[
\begin{array}{cc}
.92 & .51 \\
.39 & .86
\end{array}
\right]
\]

To quantify the estimates of the sources, we choose a distance that is invariant under a scale transformation (since the sources are estimated up to a scale factor):
\beq
\delta\pth{s_1(t),s_2(t)} = 1 - \frac{<s_1(t),s_2(t)>}{\|s_1\|.\|s_2\|}
\eeq
where $<.,.>$ and $\|.\|$ stand for the scalar product and the $L^2$ norm respectively. $\delta$ is positif and upper bounded by 1.

In order to quantify the estimates of the mixing matrix, we measure the observation distance defined by:
\beq
\delta_A = \frac{1}{m}\sum_i^m\delta\pth{\hat{\xb}_i(t),\xb_i(t)} 
\eeq
where $\hat{\xb}(t) = \hat{\Ab}\sb(t)$ and $\xb(t) = \Ab\sb(t)$.
In the simulated example, $\delta_A = 2.28 \times 10^{-4}$.

\section{Non linear MCMC Implementation}
\label{Fast.Algo}
The implementation of the proposed MCMC algorithm is modified by making use of the non linear approximation of the wavelet transfrorm:
\beq
f_M[n] = \sum_{\{j,k\}\in I_M} <f,\psi_{j,k}> \psi_{j,k}[n]
\eeq
where $I_M$ corresponds to the largest coefficients, and $f_M[n]$ is the non linear approximation of $f[n]$ by the $M$ largest coefficients. It is implemented by applying some non linear function to the wavelet coefficients of the form:
\begin{equation*}
T(<f,\psi_{j,k}>) = \left\{
\begin{array}{ll}
<f,\psi_{j,k}> & \textrm{for}\quad |<f,\psi_{j,k}>| \geq \chi \\
0 & \textrm{elsewhere}
\end{array}
\right.
\end{equation*}
known as hard thresholding. We define equivalently the soft thresholding by:
\begin{equation*}
T(<f,\psi_{j,k}>) = \left\{
\begin{array}{ll}
<f,\psi_{j,k}> - \chi & \textrm{for}\quad |<f,\psi_{j,k}>| \geq \chi \\
0 & \textrm{elsewhere}
\end{array}
\right.
\end{equation*}

In step 1 of the MCMC algorithm, the hidden variable $\zb^j$ is sampled from the \post probability $P\pth{\zb^j|.}$. The non linear approximation procedure consists then of \textit{sampling only the coefficients that are large} (in a high energy state), that corresponds to $\zb^j \in H$:
\begin{eqnarray*}
1.\quad\zb^j~\sim~P\Pth{\zb^j}{\Wbx,\theb} &=& \int_{w_s} P\Pth{\zb^j,\Wbs}{\Wbx,\theb}\\
&=&\pi\pth{\zb^j}\int_{w_s} P\Pth{\Wbx}{\Ab\Wbs,\Re}\pi\Pth{\Wbs}{\zb^j,\Rt}\\
&=& [~post_L,post_H~]^n \quad\Rightarrow \zb^j \in \{L,H\}^n
\end{eqnarray*}
the sampling of the sources coefficients with a thresholding procedure is then:
\[
2.\quad\left\{
\begin{array}{ccl}
\Ws \big| \left(z^j = L\right) &=& 0\\
\smallskip\\
\Ws \big| \left(z^j = H\right) &=& \Nc\Pth{\Ws}{0,\tau_H}
\end{array}
\right.
\]
We point out to the fact that we do not have to specify the threshold $\chi$, which is a hard task by itself, it is automatically set by the classification of the coefficients into \textit{Low energy} coefficients and \textit{High energy} coefficients. This additional procedure allows to have estimates free from any residual noise, as will be seen in the simulations, and the whole algorithm could be described as separation/denoising algorithm.

\begin{figure}[!htb]
\begin{tabular}{cc}
\includegraphics[width=50mm,height=50mm]{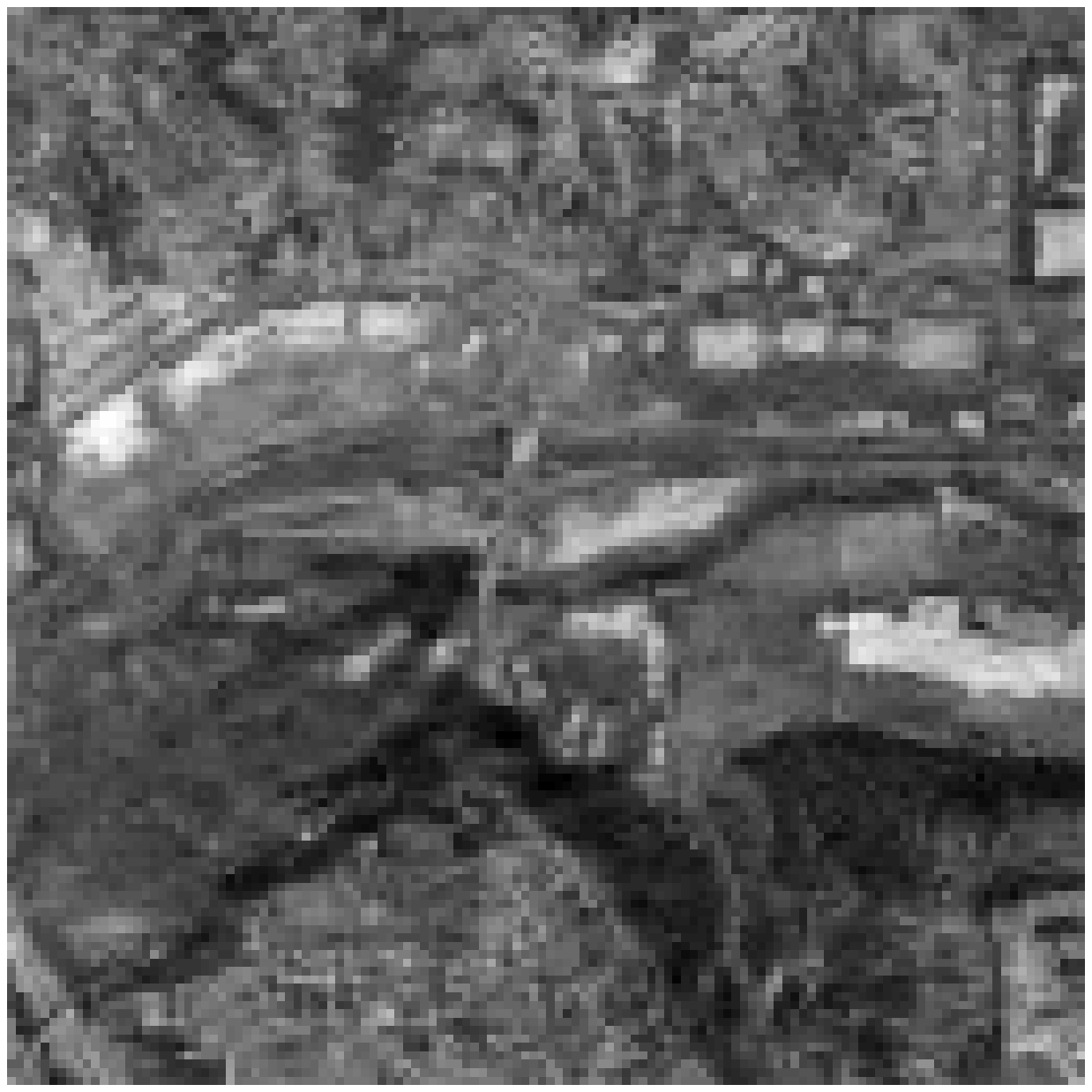} &
\includegraphics[width=50mm,height=50mm]{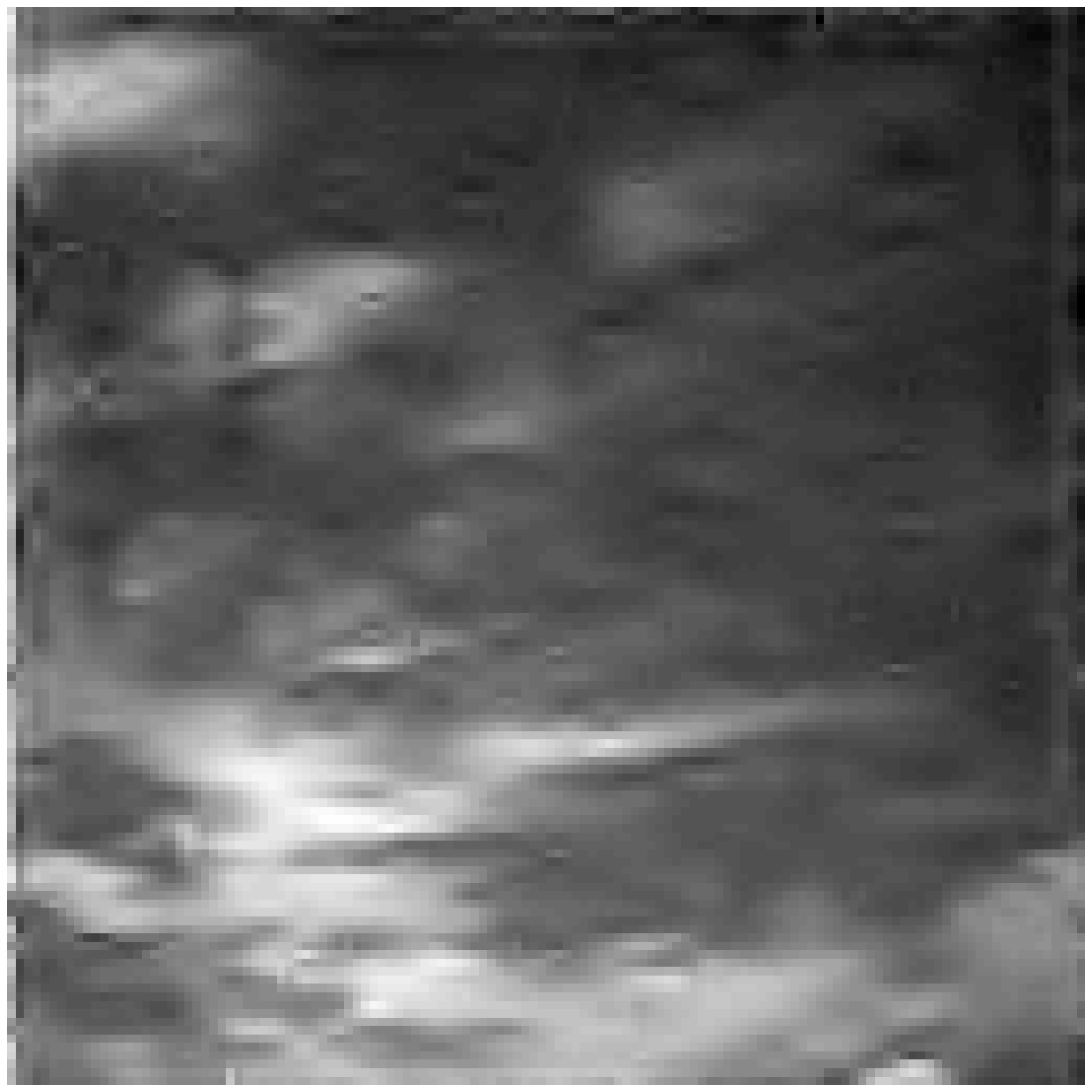}\\
$\delta = .06 $ & $\delta = .03$ \\ 
\end{tabular}
\caption{estimated sources (non linear MCMC algorithm)}
\label{Fast.Est}
\end{figure}

\begin{figure}[!htb]
\begin{tabular}{cc}
a. \includegraphics[height=25mm,width=65mm]{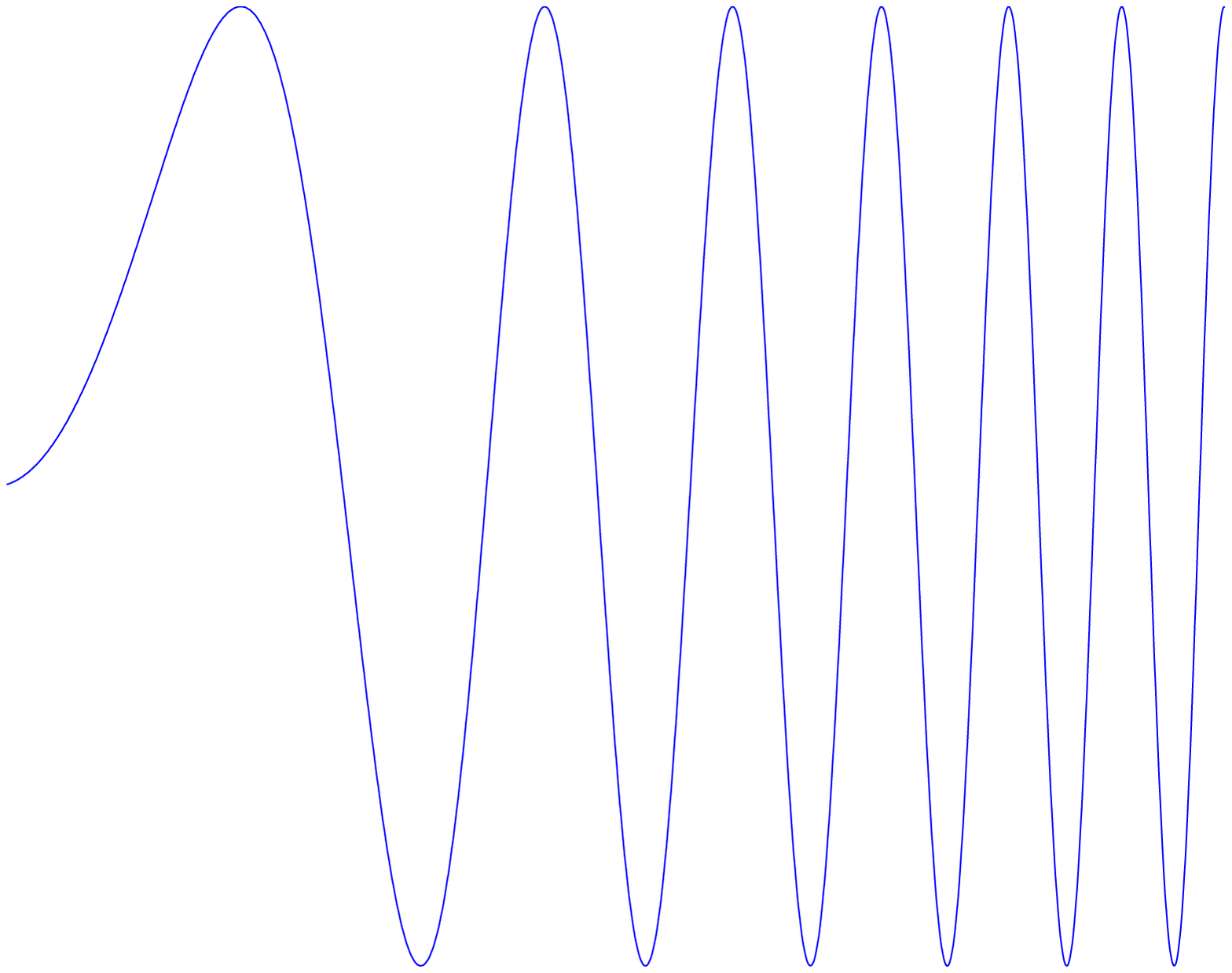} &
\includegraphics[height=25mm,width=65mm]{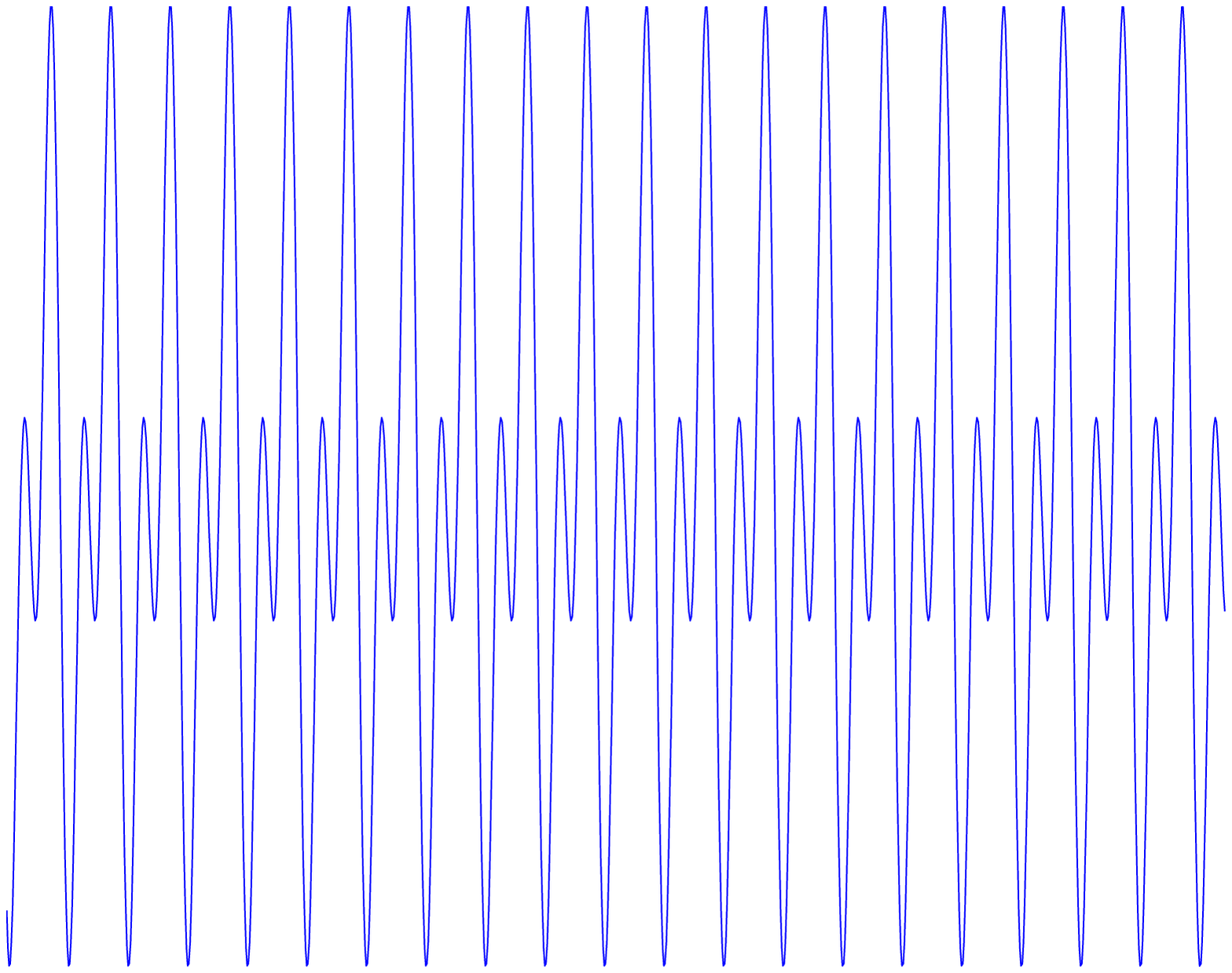} \\
b. \includegraphics[height=25mm,width=65mm]{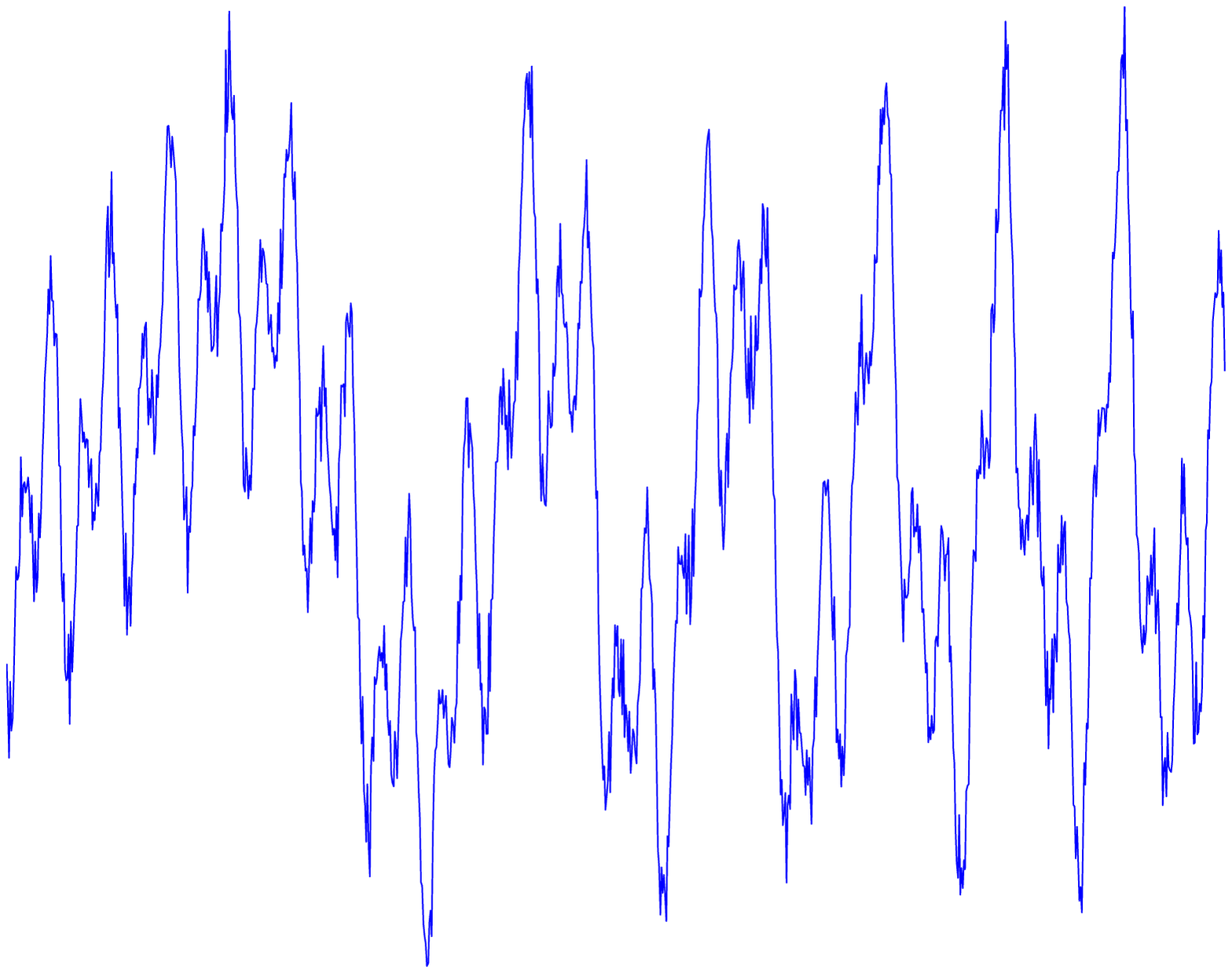} &
\includegraphics[height=25mm,width=65mm]{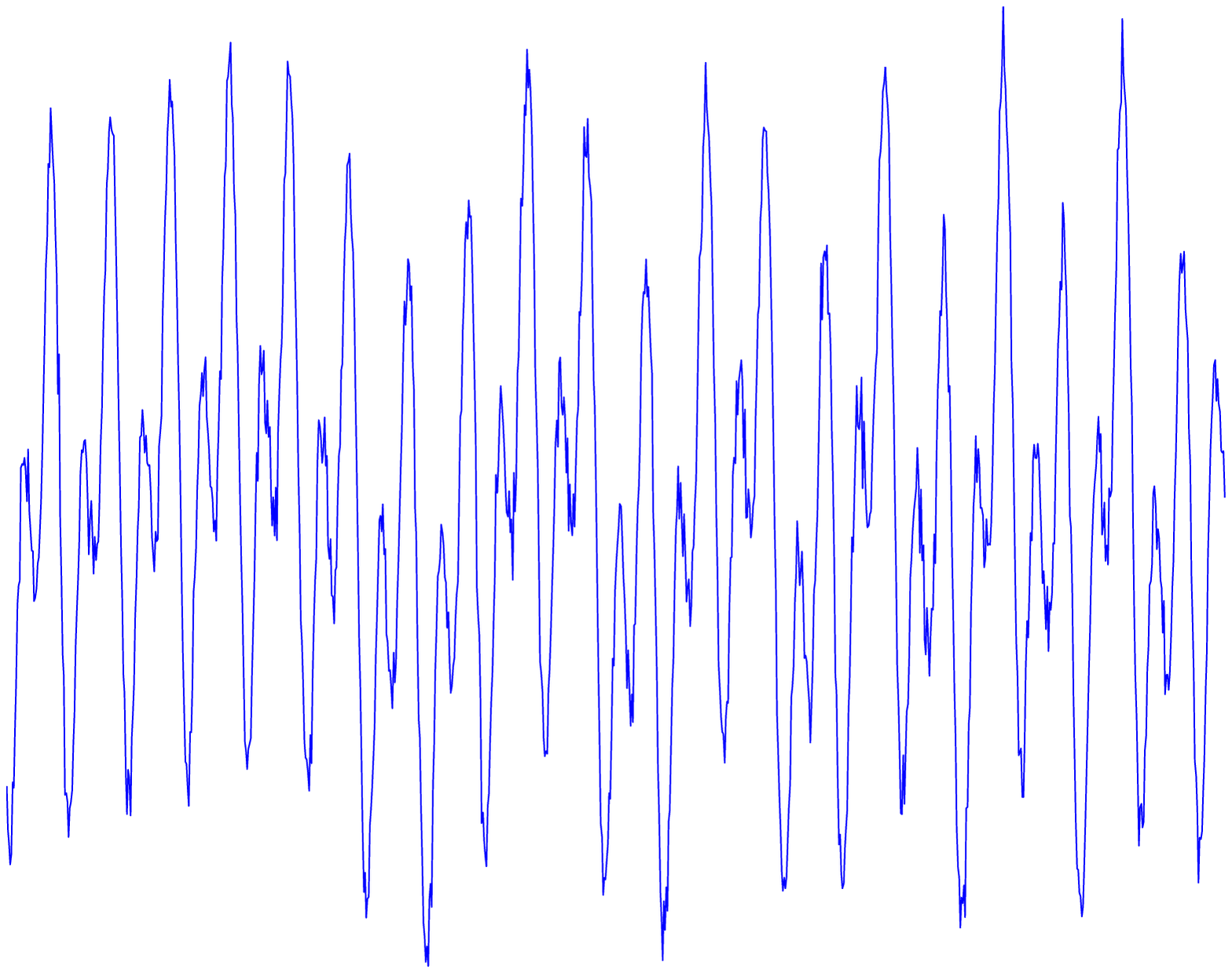} \\
$\delta = .21 $   & $\delta = .06 $   \\
c. \includegraphics[height=25mm,width=65mm]{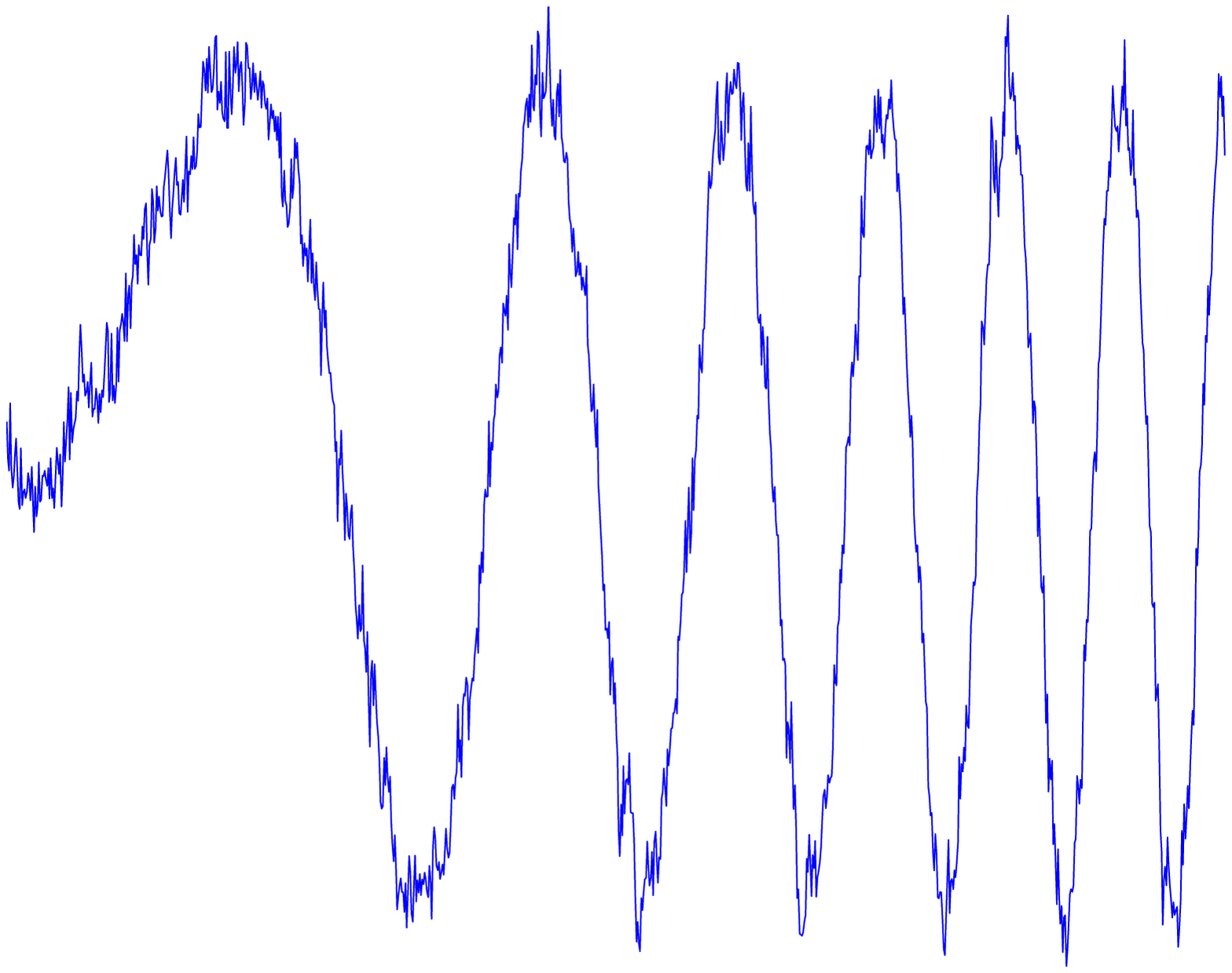} &
\includegraphics[height=25mm,width=65mm]{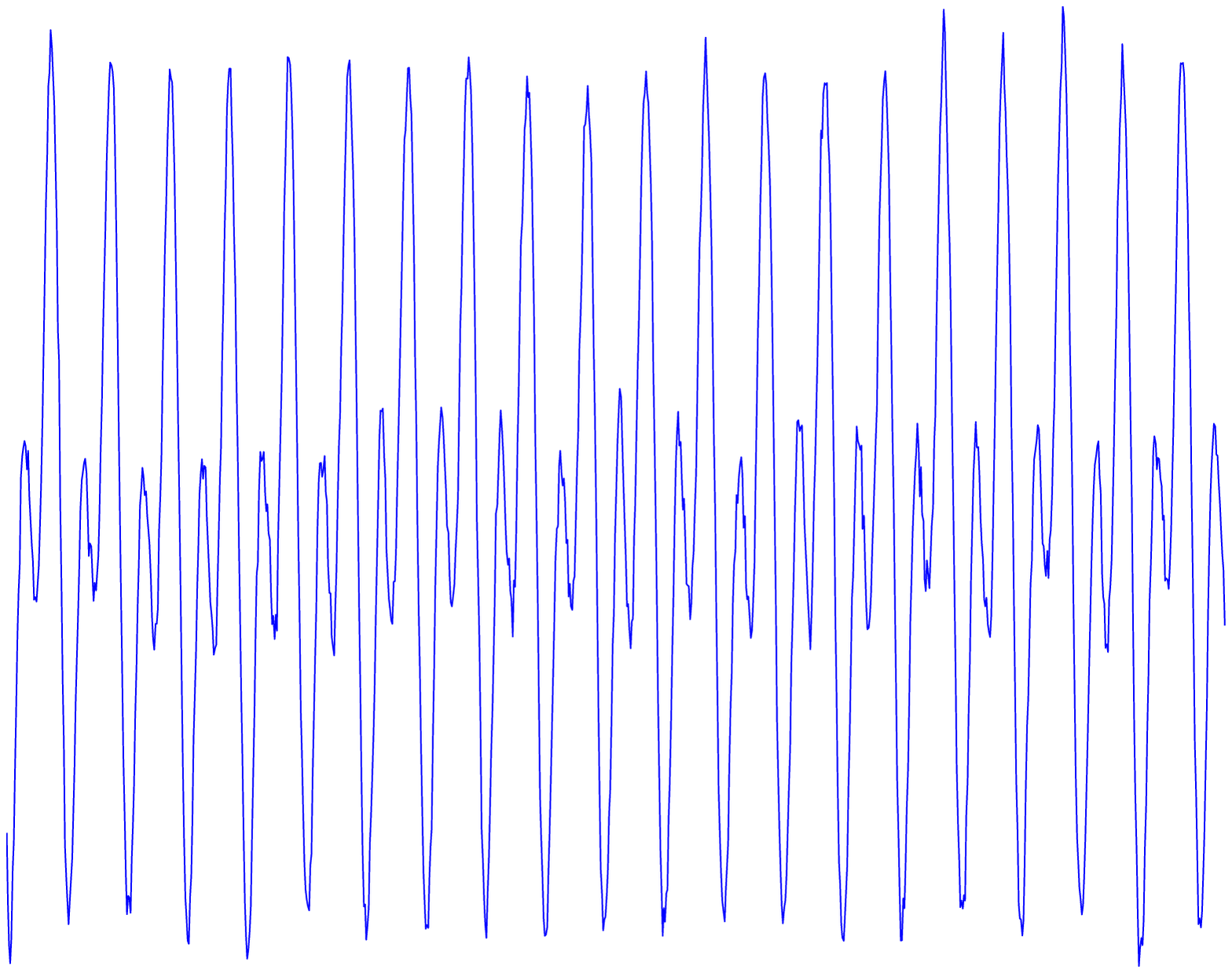} \\
$\delta = .007 $   & $\delta = .004 $   \\
d. \includegraphics[height=25mm,width=65mm]{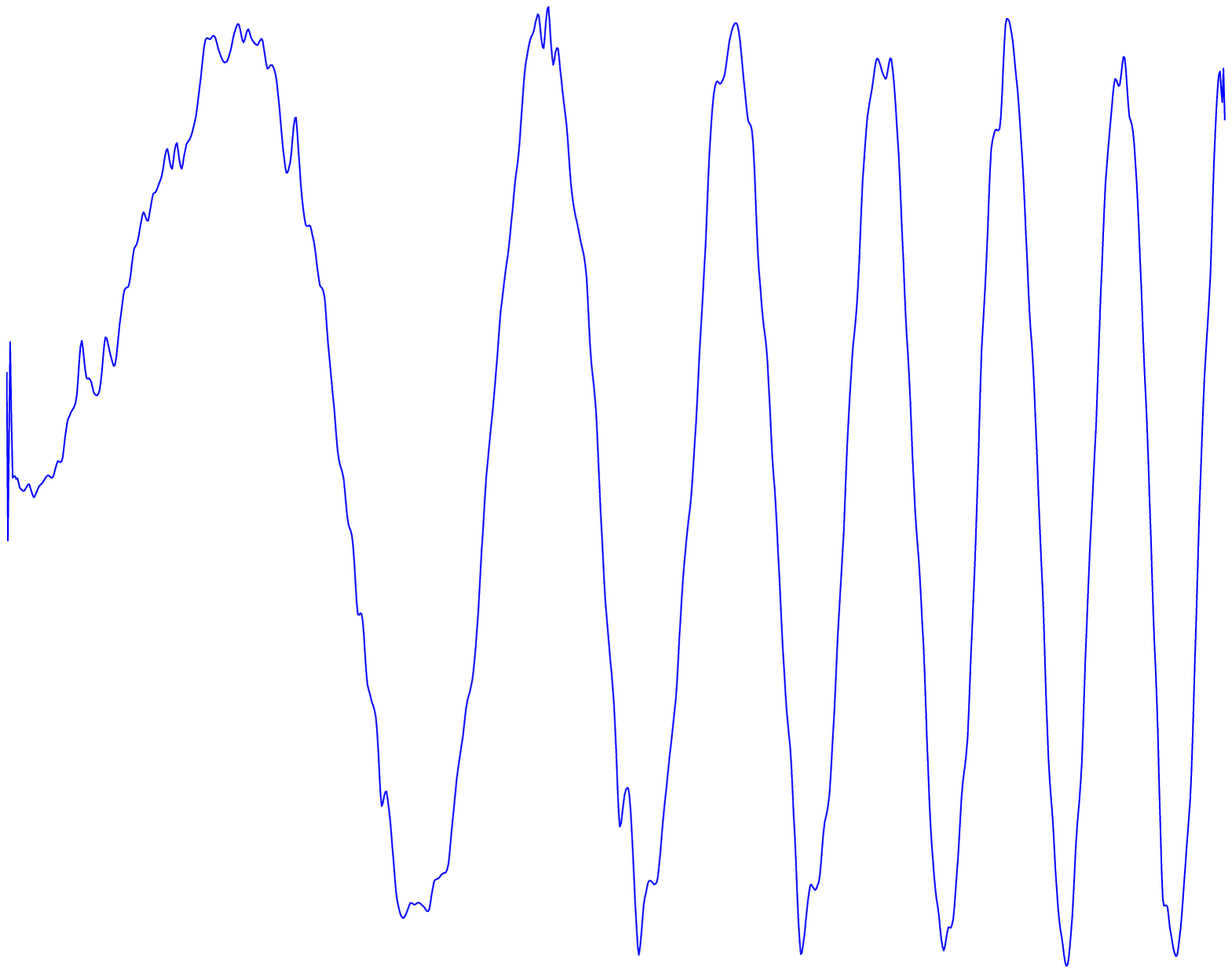} &
\includegraphics[height=25mm,width=65mm]{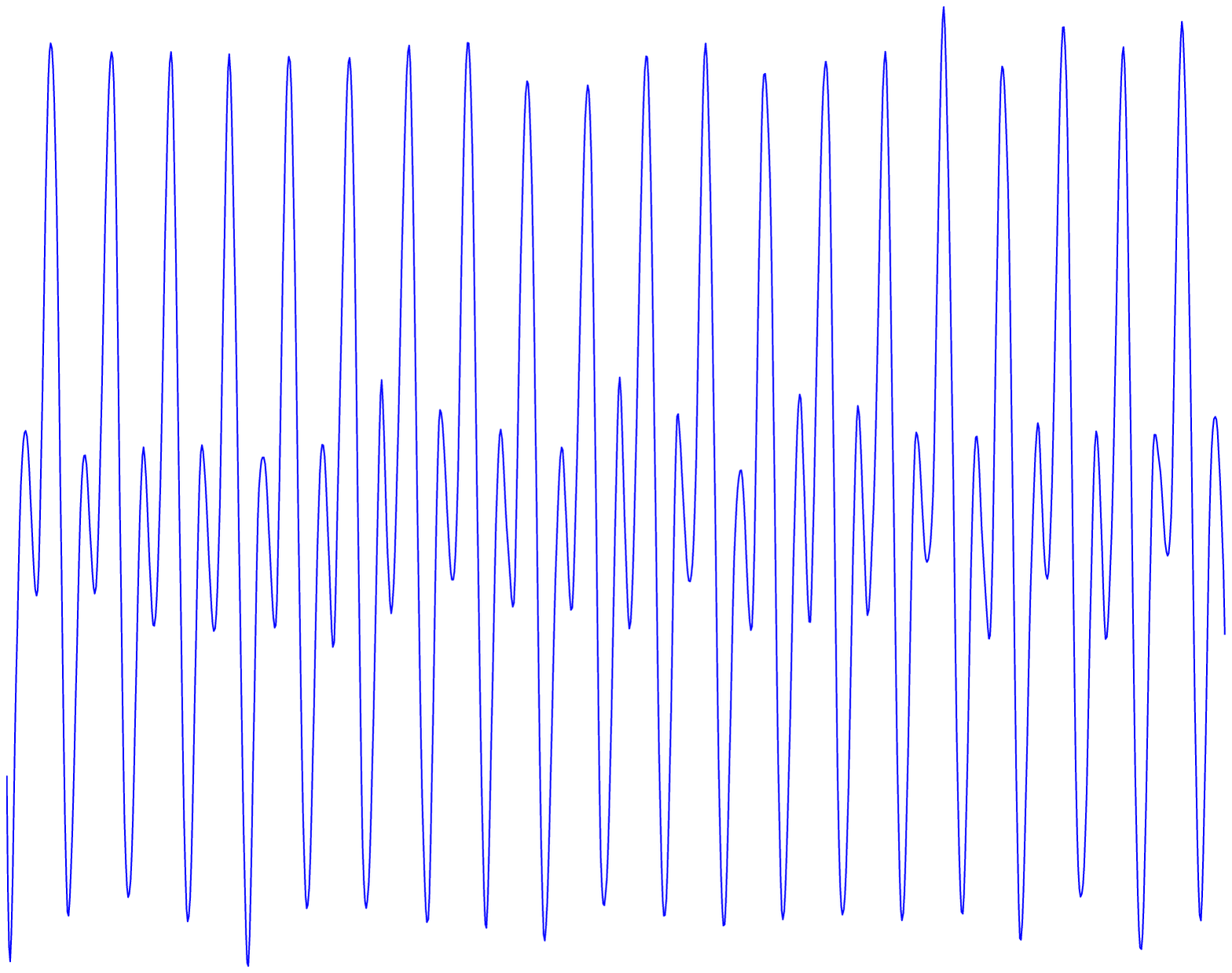} \\
$\delta = .004 $   & $\delta = .004 $
\end{tabular}
\caption{simulation results on time series signals: a. originals sources, b. mixed data (SNR = 20dB), c. estimated sources without thresholding, d. estimated sources with thresholding.}
\label{1D.Simul}
\end{figure}

\begin{figure}[!htb]
\begin{tabular}{cc}
a. \includegraphics[width=50mm,height=50mm]{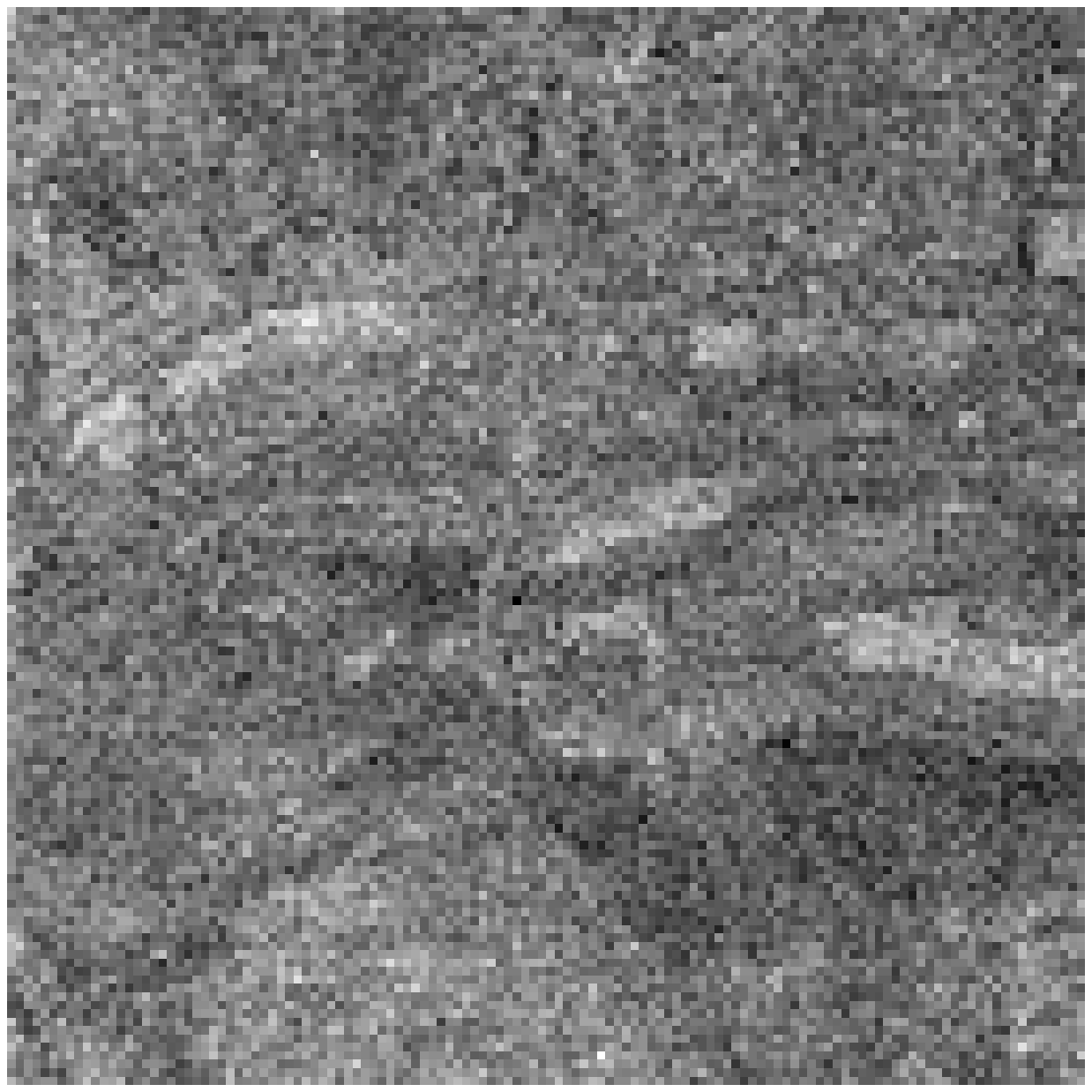} &
\includegraphics[width=50mm,height=50mm]{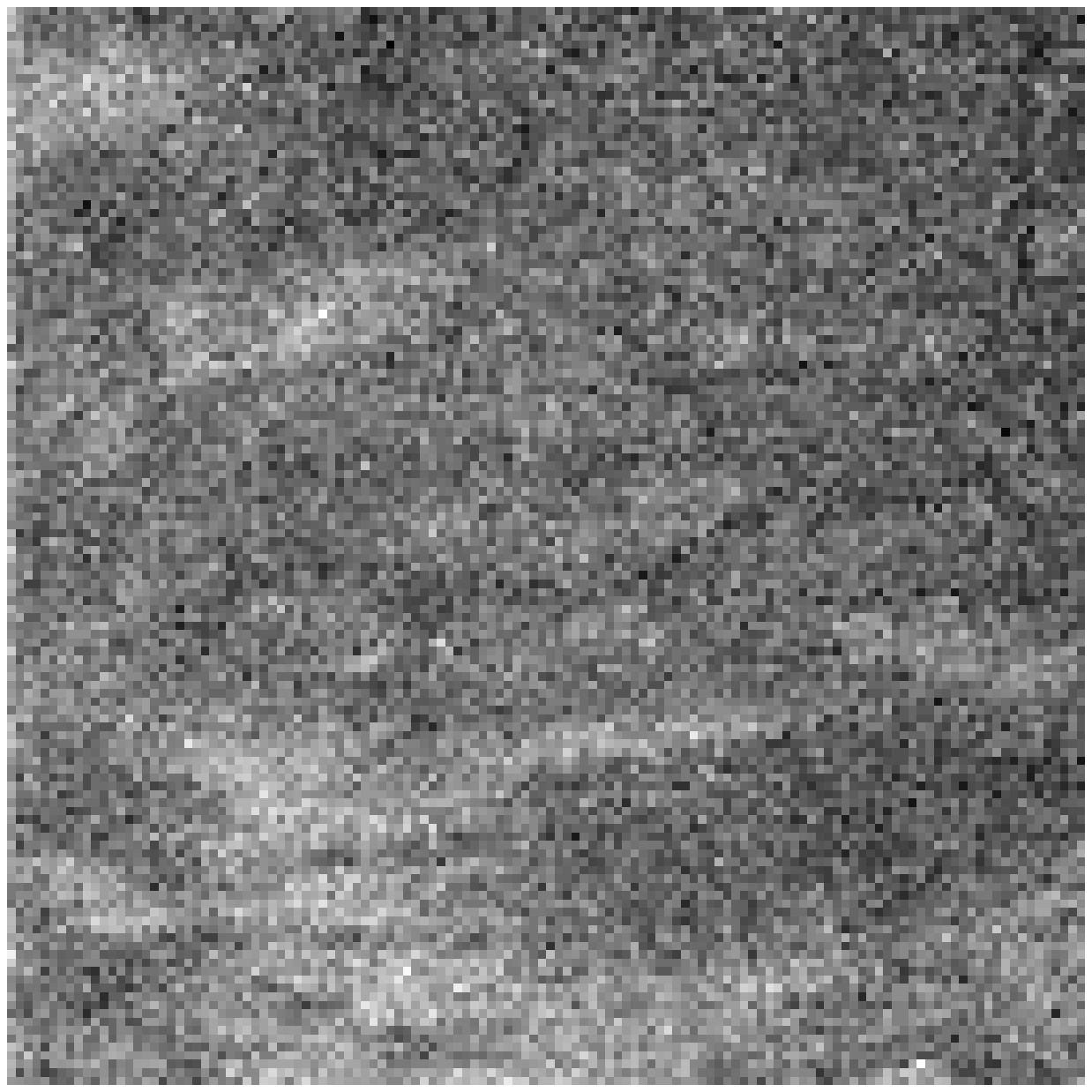}\\
$\delta = .36 $   & $\delta = .39 $   \\
b. \includegraphics[width=50mm,height=50mm]{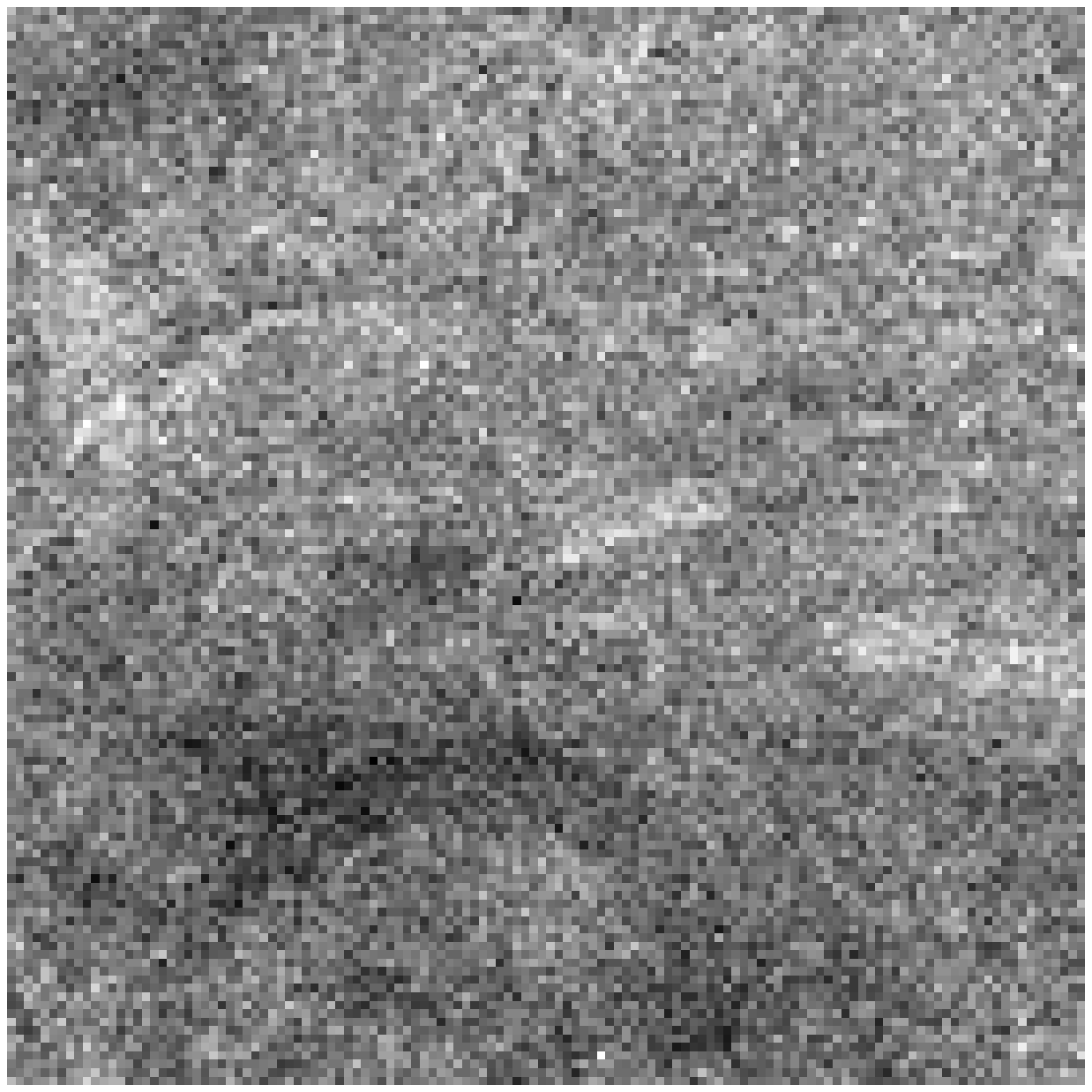} &
\includegraphics[width=50mm,height=50mm]{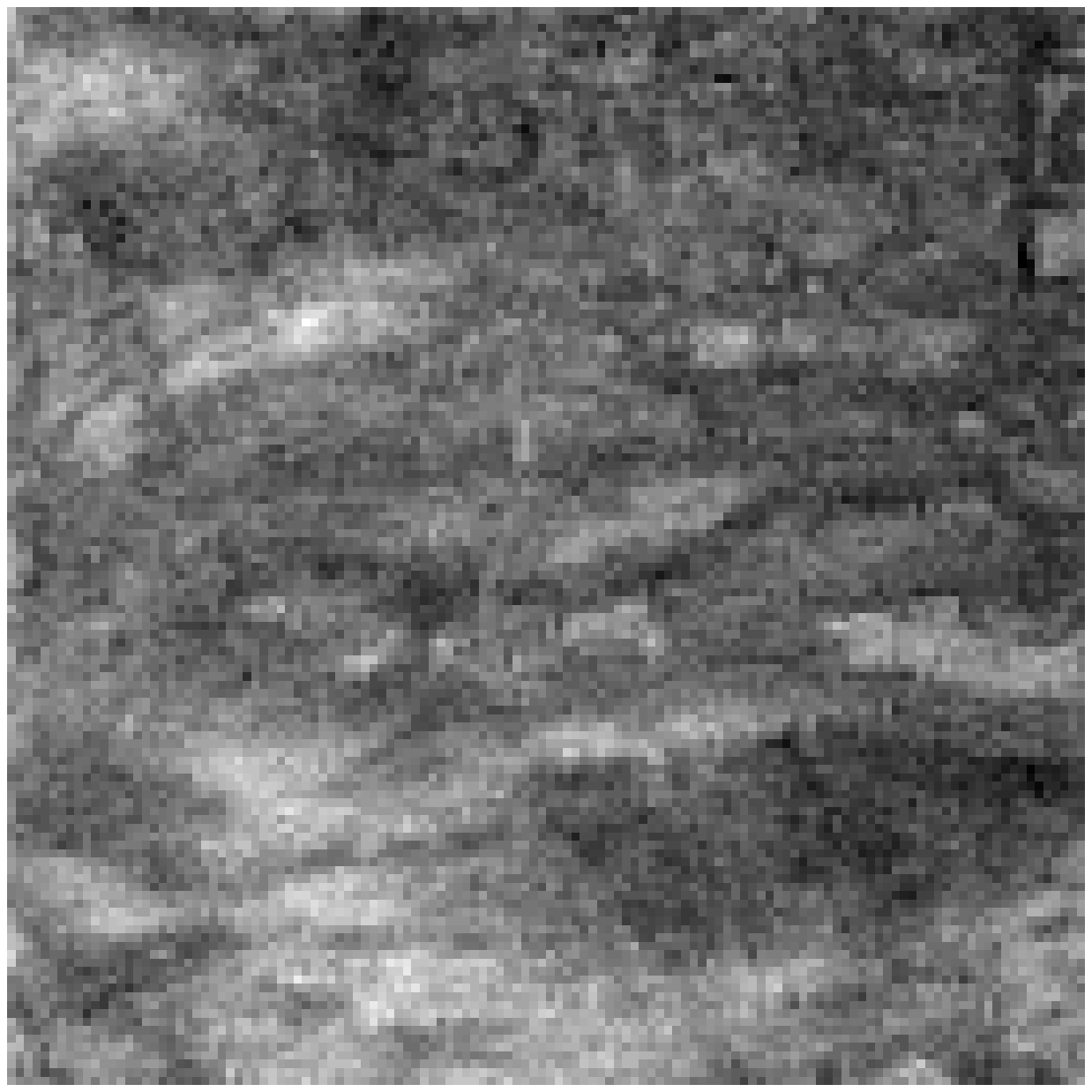}\\
$\delta = .47 $   & $\delta = .27 $   \\
c. \includegraphics[width=50mm,height=50mm]{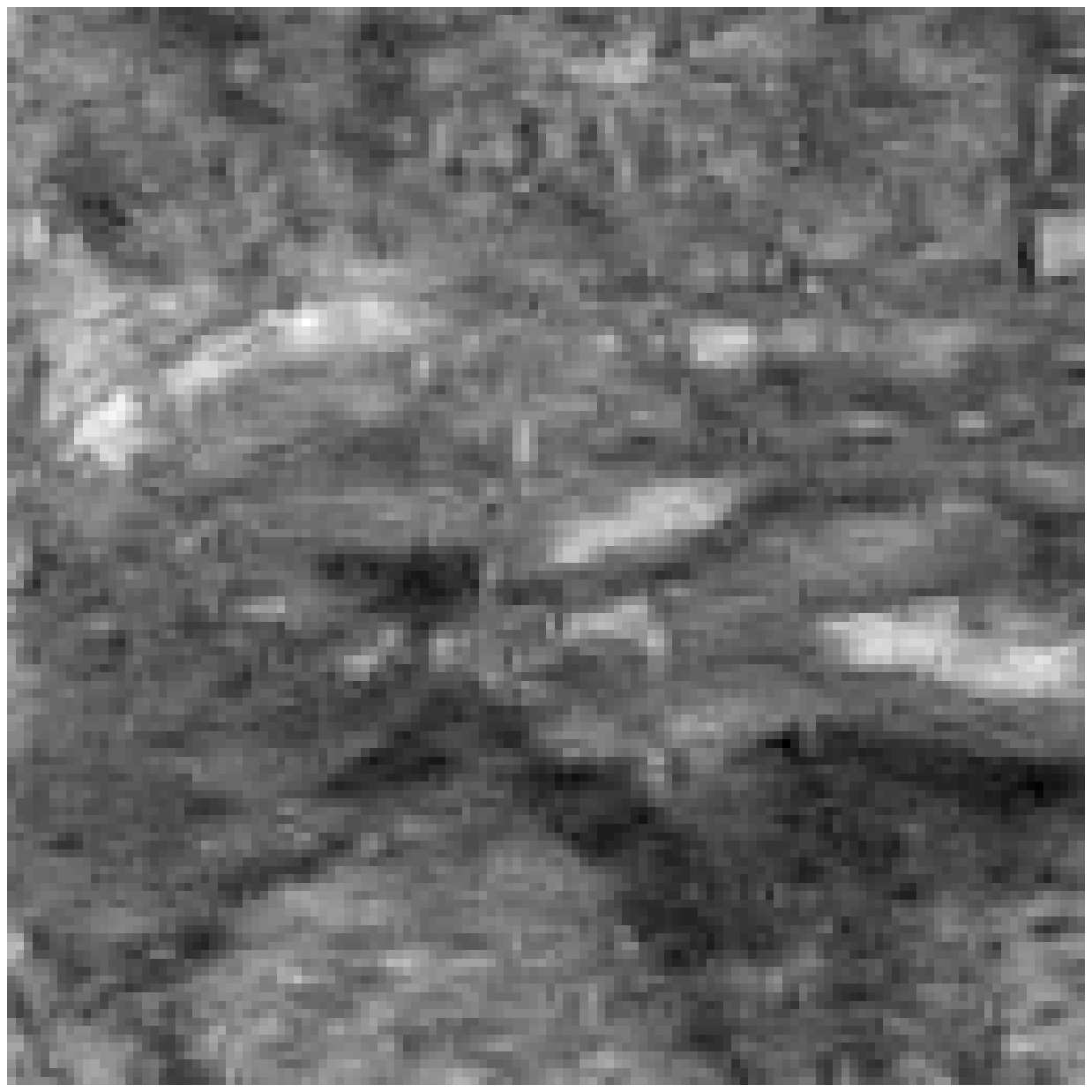} &
\includegraphics[width=50mm,height=50mm]{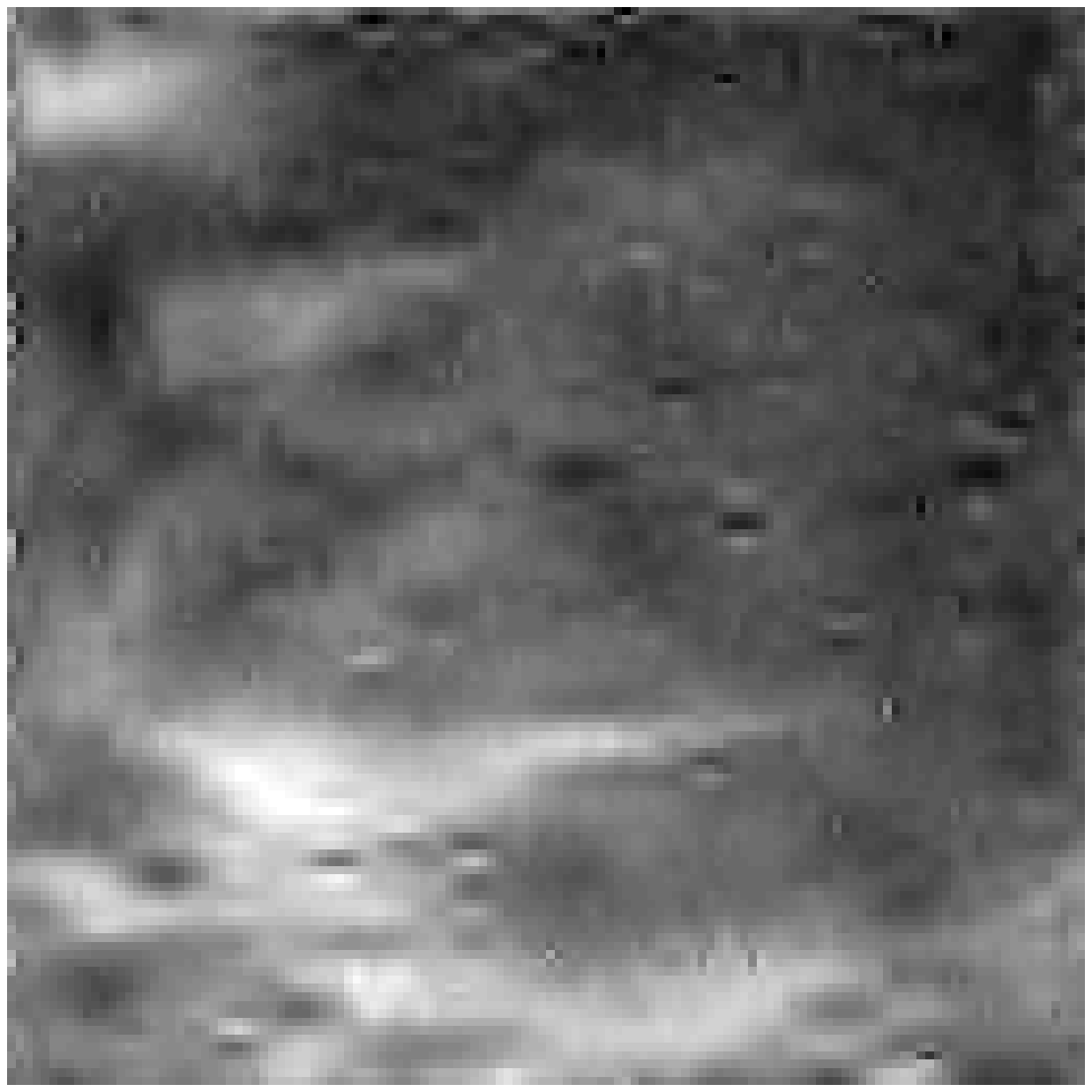}\\
$\delta = .18 $   & $\delta = .08 $ 
\end{tabular}
\caption{Example 2: a. mixed noisy images (SNR $\approx$ 10dB), b. results of separation without thresholding, c. results of separation with application of the thresholding.}
\label{Example2}
\end{figure}

The non linear step has been applied to the same data set, the estimation results are presented in figure \ref{Fast.Est} and the estimated mixing matrix is:
\[
\hat{\Ab} = \left[
\begin{array}{cc}
.89 & .51 \\
.46 & .86
\end{array}
\right]
\]
and the observation distance $\delta_A = 9.16\times 10^{-4}$.\\
The algorithm has been tested on 1D signals and the results presented in figure \ref{1D.Simul} show the effect of the non linear MCMC implementation on denoising the estimates. In figure \ref{Example2}, a second example is presented where the additional information brought by this non linear procedure is very apparent in the sense that it helps for separating the sources in a very noisy environment.

\section{Summary and Perspectives}
In this work we presented a Bayesian aproach to blind source separation in the wavelet domain. The main interest to try to solve the problem in the wavelet domain is to be able to use a simpler probabilistic model for the sources i.e. a two component Gaussian mixture model with a total of three parameters as opposed to a $3L-1$ parametric model in the direct model with $L$ undetermined. Indeed, the interpretation of the mixture model as a heirarchical hidden variable model gives us the ability to apply some automatic thresholding rule to the wavelet coefficients. Finally, we showed some performances of the proposed method on simulated data.

\medskip
\noindent Concerning our perspectives, we follow essentially these directions:

i) a quad tree Markovian modeling of the wavelet coefficients to account for inter-scale correlation.

ii) an adaptative basis selection criteria to improve the thresholding procedure.


\begin{figure}[!htb]
\begin{tabular}{ccc}
a. \includegraphics[height=40mm,width=40mm]{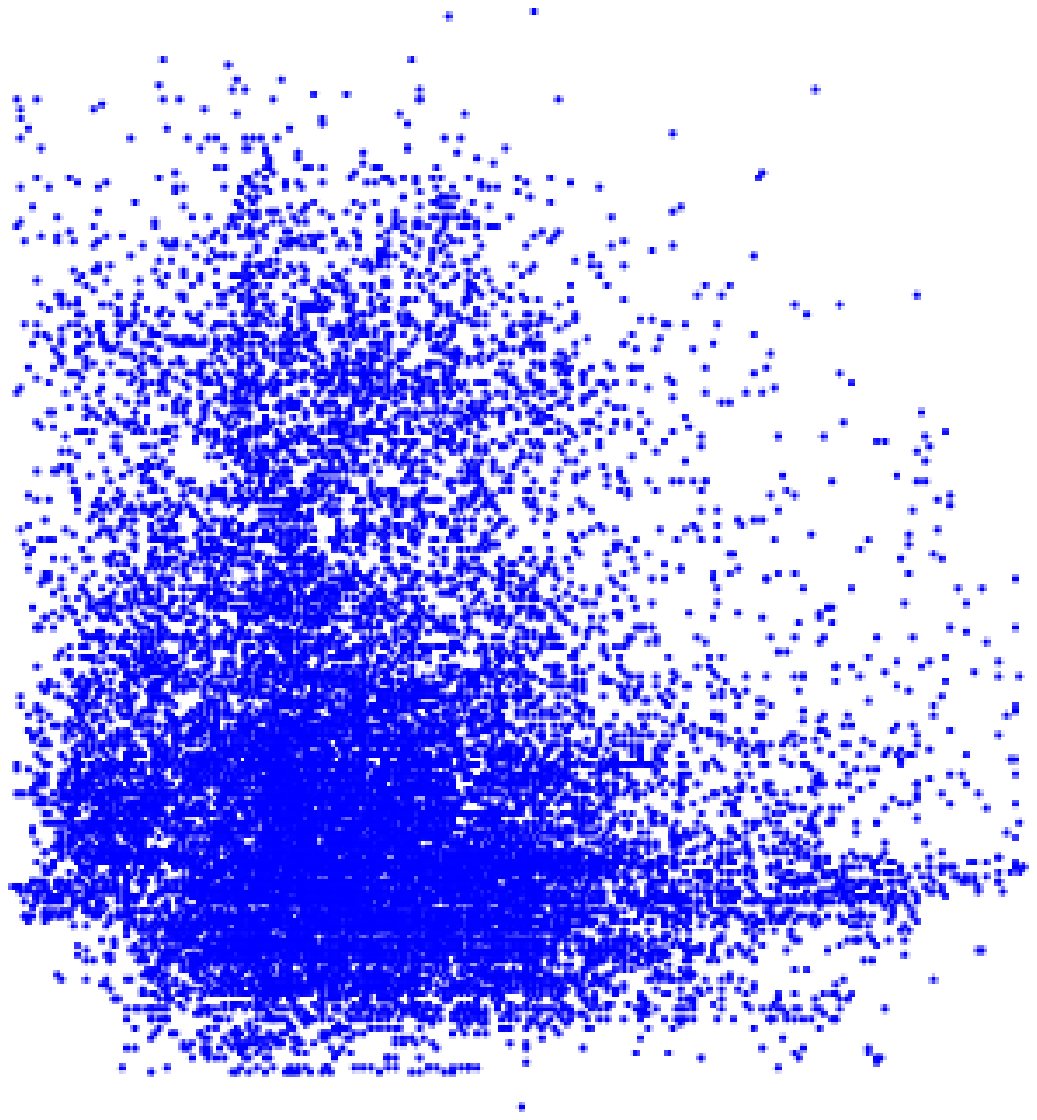} &
b. \includegraphics[height=40mm,width=40mm]{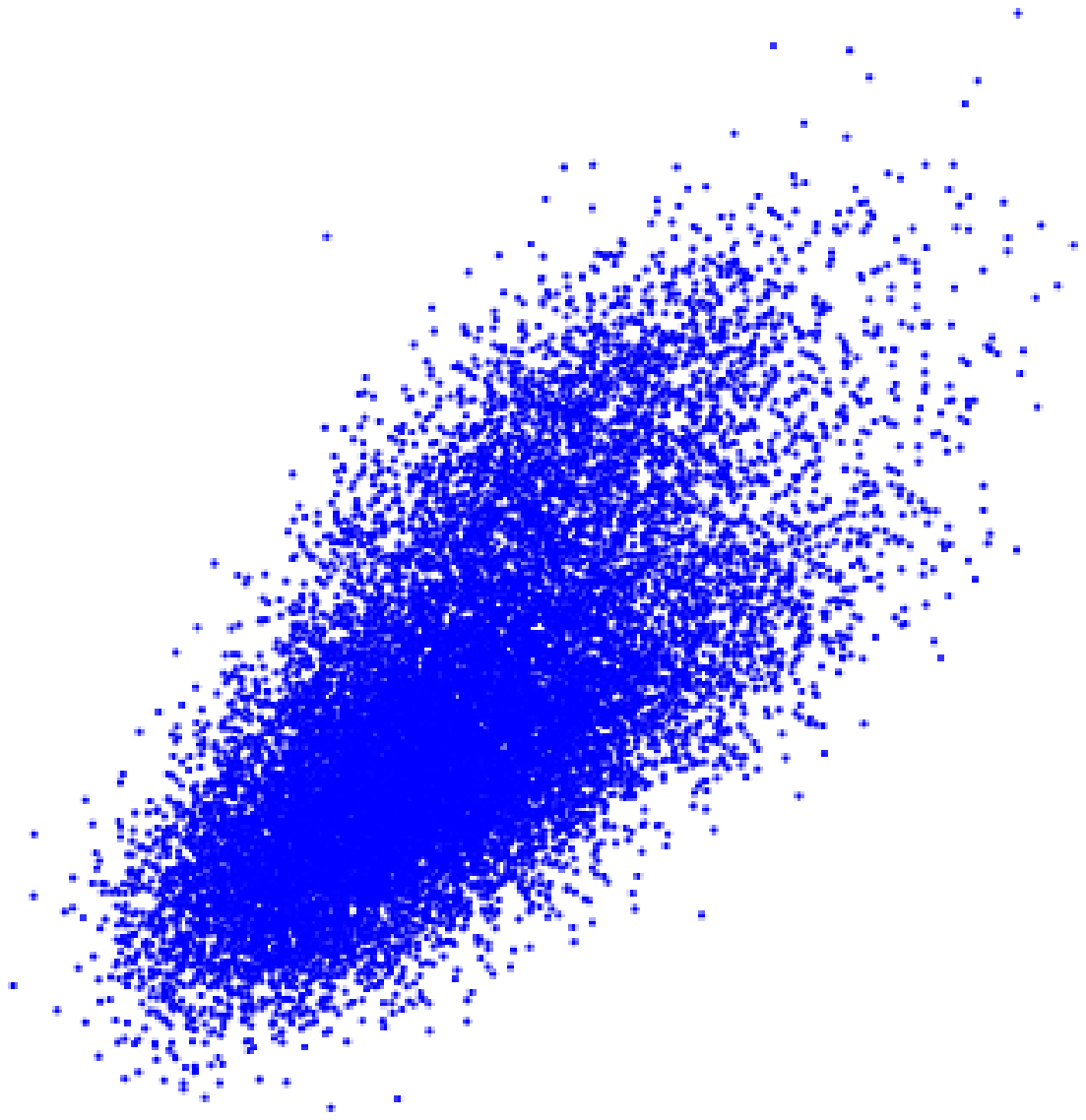} &
c. \includegraphics[height=40mm,width=40mm]{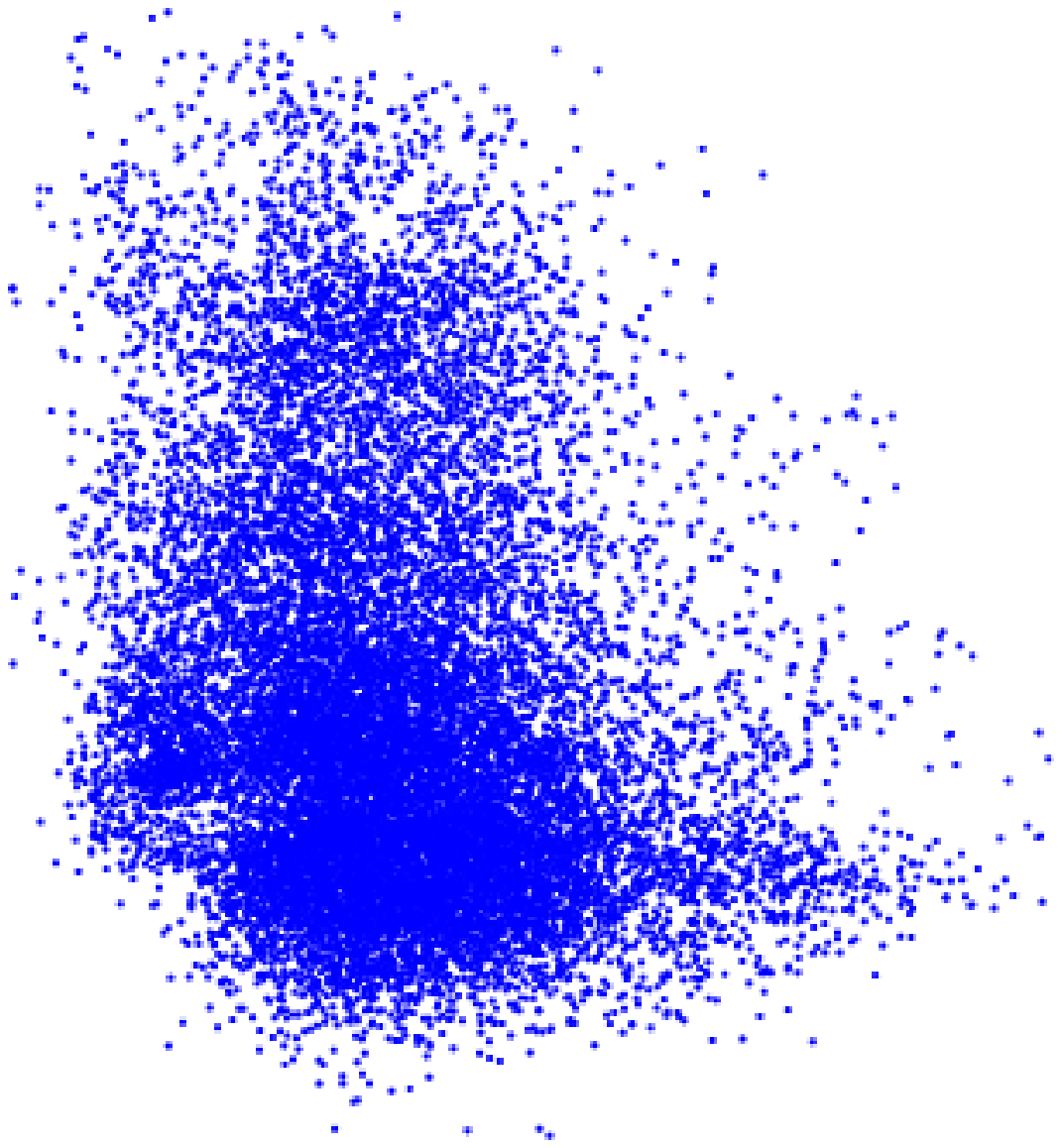}
\end{tabular}
\caption{scattering plots of : a. originals sources, b. mixed data, c. estimated sources.}
\end{figure}

\bibliographystyle{plain}
\bibliography{biben,revueabr,BaseAZ,gpipubli}


\end{document}